\documentclass[a4paper,fleqn]{cas-dc}

\usepackage[numbers]{natbib}

\usepackage{subcaption}

\usepackage{enumitem}

\usepackage{siunitx}

\usepackage{threeparttable}

\usepackage{array}
\usepackage{booktabs}
\usepackage[table,xcdraw]{xcolor}

\PassOptionsToPackage{normalem}{ulem}
\usepackage{ulem}

\def\tsc#1{\csdef{#1}{\textsc{\lowercase{#1}}\xspace}}
\tsc{WGM}
\tsc{QE}

\begin{document}
\let\WriteBookmarks\relax
\def\floatpagepagefraction{1}
\def\textpagefraction{.001}

\shorttitle{Sequential topology optimization}

\shortauthors{O. Jezek et~al.}

\title [mode = title]{Sequential topology optimization: SIMP initialization for level-set boundary refinement}

\tnotemark[1] 

%

\author[1,2]{Ondřej Ježek}[orcid=0000-0002-5380-442X]
\ead{ondrej.jezek@fs.cvut.cz}

\author[1]{Ján Kopačka}[orcid=0000-0002-2975-8347]
\ead{kopacka@it.cas.cz}

\author[1]{Martin Isoz}[orcid=0000-0002-5862-2561]
\ead{isozm@it.cas.cz}

\author[1]{Dušan Gabriel}[orcid=0000-0002-7691-2191]
\ead{gabriel@it.cas.cz}

\affiliation[1]{organization={Institute of Thermomechanics, Czech Academy of Sciences},
            addressline={Dolejškova 1402/5}, 
            postcode={182 00, Praha 8}, 
            state={Czech Republic}}

\affiliation[2]{organization={Faculty of Mechanical Engineering, Czech Technical University in Prague},
            addressline={Technická 4}, 
            postcode={160 00, Praha 6}, 
            state={Czech Republic}}

\begin{abstract}
Density-based topology optimization methods such as SIMP enable efficient topological exploration but produce diffuse material boundaries that require interpretation before manufacturing. Level-set methods maintain sharp interfaces but are sensitive to the initial design. This paper presents a sequential framework that addresses these complementary limitations through a signed distance function (SDF)-based geometry transfer, formulated for three-dimensional meshes. The SIMP density distribution is converted into an SDF that initializes subsequent level-set boundary refinement. From the level-set perspective, the SIMP-derived initialization mitigates sensitivity to the initial design. From the SIMP perspective, the level-set stage acts as optimization-driven post-processing that produces manufacturing-ready boundaries. Validation on three-dimensional cantilever and MBB benchmarks demonstrates compliance comparable to standalone level-set optimization, with up to 4.6$\times$ wall-clock speedup on the cantilever case. The full implementation is released under an open-source license to support reproducibility.
\end{abstract}

%



\maketitle

\section{Introduction}\label{sect:introduction}

\noindent The availability of additive manufacturing and other modern production technologies has made topology optimization an increasingly essential tool for efficient structural design~\cite{Sigmund2013, Deaton2014}. By determining optimal material distributions within a design domain subject to prescribed constraints, topology optimization enables engineers to discover structural configurations that would be difficult to conceive through conventional design processes. Applications span multiple industries, including aerospace, automotive, and biomedical engineering~\cite{Meng2019, Liu2018}.

In industrial practice, the Solid Isotropic Material with Penalization (SIMP) method remains the dominant approach~\cite{Deaton2014, Sigmund2013} due to its mathematical simplicity, straightforward sensitivity analysis, and reliable convergence properties~\cite{Bendsoe1999}. The fixed-grid discretization integrates directly with finite element solvers, and density-based formulations are implemented in all major CAE packages offering topology optimization capabilities, including, among others, Altair OptiStruct, ANSYS, and TOSCA Structure~\cite{OptiStruct2026, ANSYS2025, TOSCA2025}.

These methods produce element-wise constant density distributions with inherent geometric artifacts: stair-stepped interfaces reflecting the underlying mesh discretization, and intermediate density values representing neither solid nor void. These characteristics require interpretation before results can be translated into manufacturable designs~\cite{Subedi2020}. As manufacturing capabilities have advanced, research attention has increasingly shifted toward producing high-quality geometries suitable for direct fabrication, driving developments in level-set optimization and geometric post-processing~\cite{Subedi2020, Li2022, Wegert2025}.

Level-set methods address the boundary definition limitation by representing the structural domain implicitly as the zero-level set of a continuous function~\cite{Wang2003, Allaire2004}. This implicit representation enables natural handling of topological changes such as hole merging and splitting without explicit boundary parameterization~\cite{Wang2006RBF, VanDijk2013}. Additionally, sharp and well-defined interfaces are maintained throughout the optimization process. The resulting geometries require minimal post-processing for manufacturing applications~\cite{Deaton2014}.

However, level-set approaches face specific challenges. Convergence is typically slower than density-based methods because design updates occur only in the vicinity of existing interfaces~\cite{VanDijk2013}. Standard formulations also exhibit sensitivity to initial design specification, as they cannot nucleate new holes without dedicated mechanisms~\cite{Allaire2004}. Topological derivative-based approaches address this limitation~\cite{Allaire2005, Burger2004}, but require careful parameter tuning and increase implementation complexity~\cite{VanDijk2013}. Alternative strategies include parametric level-set methods~\cite{Wang2006RBF}, reaction-diffusion formulations~\cite{Yamada2010}, and density-informed hole seeding that couples density fields with level-set evolution~\cite{Barrera2020}. More recently, unfitted finite element approaches have demonstrated enhanced boundary resolution~\cite{Wegert2025}. Despite these advances, level-set methods remain less widely adopted in commercial software relative to density-based approaches~\cite{Meng2019}.

The preceding discussion reveals complementary characteristics of the two methodologies. SIMP enables simultaneous material redistribution across the entire design domain, allowing topological changes without dedicated nucleation mechanisms. In contrast, level-set methods confine design updates to the boundary vicinity but maintain geometrically precise interfaces. These characteristics suggest a natural division of labor: density-based optimization for rapid topology exploration, level-set refinement for boundary quality improvement. The use of density-based results to initialize level-set optimization has been noted as a viable strategy~\cite{Allaire2004}, directly addressing the initialization sensitivity inherent to level-set methods. Sequential frameworks combining different optimization approaches have demonstrated practical benefits in various contexts~\cite{Li2025}. A related two-stage approach combining NURBS reconstruction with shape optimization of density-based results was proposed by \cite{Tang2026}, although the framework is restricted to two-dimensional problems. The first systematic sequential SIMP-to-level-set framework, to our knowledge, was proposed by \cite{Swierstra2017, Swierstra2020}, who demonstrated post-processing of both 2D and 3D density-based results. Their formulation operates on the original SIMP grid without remeshing but relies on uniform element spacing, restricting the methodology to structured meshes.

The present work builds on the sequential concept of~\cite{Swierstra2020} but introduces an SDF-based geometry transfer formulated for arbitrary mesh types, providing a well-conditioned initialization for the level-set equation. The extraction procedure draws on the SDF construction methodology of~\cite{Jezek2026} in simplified form, as the subsequent level-set optimization inherently refines the boundary. From the SIMP perspective, the level-set stage serves as optimization-based post-processing that can improve boundary geometry. This coupling strategy was identified in~\cite{Jezek2026} as a future research direction. From the level-set perspective, the SIMP-derived initialization mitigates sensitivity to the initial design and accelerates convergence by providing a feasible, topologically informed starting configuration. The specific contributions of this work are:
\begin{enumerate}
  \item \textbf{Mesh-agnostic geometry transfer} from SIMP density fields to a level-set representation through SDF construction~\cite{Jezek2026}.
  \item \textbf{Coupling strategy} that positions level-set refinement as optimization-driven post-processing of SIMP results, producing sharp manufacturing-ready boundaries without purely geometric smoothing.
  \item \textbf{Validation} on 3D cantilever and MBB benchmarks with open-source implementation, demonstrating up to $4.6\times$ wall-clock speedup versus porous-initialization level-set baselines.
\end{enumerate}

The paper is organized as follows: Section~\ref{sect:background} provides the theoretical foundations of SIMP and level-set topology optimization methods. The proposed sequential optimization framework, including the pipeline overview and the algorithmic details of each stage, is presented in Section~\ref{sect:methodology}. Section~\ref{sect:num_tests} demonstrates the effectiveness of the methodology through numerical experiments and comparative analysis. Conclusions and recommendations for future work are given in Section~\ref{sect:conclusion}.
The open-source implementation\footnote{\url{https://github.com/jezekon/2026-Jezek-SeqTopOpt}} 
is described in Appendix~\ref{app:data}.

\section{Background}\label{sect:background}

\noindent The proposed framework builds upon two established optimization approaches: the SIMP method for density-based optimization (Section~\ref{sub:SIMP}) and the level-set method with implicit boundary representation (Section~\ref{sub:levelset}).

\subsection{SIMP method}\label{sub:SIMP}

\noindent The SIMP method represents material distribution within a fixed finite element mesh using continuous density variables~\cite{Bendsoe1989, Zhou1991}. Each element $e$ carries a density value $\rho_e \in [0,1]$, where $\rho_e = 0$ represents void and $\rho_e = 1$ represents solid material. The element stiffness is related to density through the modified SIMP power-law interpolation~\cite{Sigmund2007}:
\begin{equation}
    E_e(\rho_e) = E_{\min} + \rho_e^p (E_0 - E_{\min})\,,
    \label{eqn:penalization}
\end{equation}
where $E_0$ is the Young's modulus of the solid material, $E_{\min}$ is a small value preventing stiffness matrix singularity, and $p$ is the penalization exponent. The penalization exponent is typically set to $p = 3$~\cite{Bendsoe1999}. For $p > 1$, intermediate densities become mechanically inefficient, which drives the optimization toward binary material distributions.

\subsubsection{Problem formulation}

\noindent For a design domain with $N$ elements, the compliance minimization problem reads:
\begin{align}
    \min_{\boldsymbol{\rho}} \quad & J(\boldsymbol{\rho}) = \mathbf{U}^\mathsf{T} \mathbf{K}(\boldsymbol{\rho}) \mathbf{U} = \sum_e \boldsymbol{u}_e^\mathsf{T} \boldsymbol{k}_e(\rho_e) \boldsymbol{u}_e \notag\\
    \text{s.t.} \quad & \mathbf{K}(\boldsymbol{\rho}) \mathbf{U} = \mathbf{F} \label{eqn:SIMP_formulation}\\
    & V(\boldsymbol{\rho}) \leq V^* \notag\\
    & 0 \leq \rho_e \leq 1\,, \notag
\end{align}
where $\mathbf{K}$ is the global stiffness matrix assembled from element contributions $\boldsymbol{k}_e$, $\mathbf{U}$ is the global displacement vector, $\boldsymbol{u}_e$ is the element displacement vector, $\mathbf{F}$ is the external force vector, $V_e$ is the volume of element $e$, $V(\boldsymbol{\rho}) = \sum_{e=1}^{N} V_e \rho_e$ is the total material volume, and $V^*$ is the prescribed upper bound~\cite{Sigmund2001, Andreassen2011}.

\subsubsection{Sensitivity analysis}

\noindent Gradient-based solution of problem~\eqref{eqn:SIMP_formulation} requires the derivative of the objective function with respect to element densities. Because compliance is self-adjoint, direct differentiation (without an adjoint solve) yields~\cite{Sigmund2001}:
\begin{equation}
    \frac{\partial J}{\partial \rho_e} = -p \rho_e^{p-1} (E_0 - E_{\min})\, \boldsymbol{u}_e^\mathsf{T} \boldsymbol{k}_e^0 \boldsymbol{u}_e\,,
    \label{eqn:sensitivity}
\end{equation}
where $\boldsymbol{k}_e^0$ is the element stiffness matrix for unit Young's modulus. These sensitivities are used in gradient-based optimizers such as the Optimality Criteria method~\cite{Bendsoe2003} or the Method of Moving Asymptotes~\cite{Svanberg1987}.

\subsubsection{Sensitivity filtering}

\noindent Standard SIMP formulations exhibit numerical difficulties including checkerboard patterns and mesh dependency~\cite{Sigmund1998}. Sensitivity filtering addresses these issues by replacing each element sensitivity with a weighted average over neighboring elements within a filter radius $R$~\cite{Sigmund1997}. The formulation applicable to general meshes with varying element volumes $V_e$ reads~\cite{Sigmund2007}:
\begin{equation}
    \widetilde{\frac{\partial J}{\partial \rho_e}} = \frac{\displaystyle\sum_{i \in \mathcal{N}_e} \frac{ \rho_i}{V_i}\,H(\boldsymbol{x}_i)\, \frac{\partial J}{\partial \rho_i}}{\displaystyle\frac{\rho_e}{V_e} \sum_{i \in \mathcal{N}_e} H(\boldsymbol{x}_i)}\,,\,\, H(\boldsymbol{x}_i) = \max(0,\, R - \|\boldsymbol{x}_e - \boldsymbol{x}_i\|)\,,
    \label{eqn:sensitivity_filter}
\end{equation}
where $\boldsymbol{x}_e$ represents the element centroid, $\mathcal{N}_e$ denotes the set of elements whose centroids lie within distance $R$ from $\boldsymbol{x}_e$, and $H(\boldsymbol{x}_i)$ is a linearly decaying weight function. The filtered sensitivities from~\eqref{eqn:sensitivity_filter} replace the original values from~\eqref{eqn:sensitivity} in the subsequent optimization update. The filter radius enforces mesh independence by preventing structural features below a certain size~\cite{Sigmund1997, Sigmund2001}. Alternative regularization approaches include density filtering~\cite{Bruns2001, Bourdin2001} and projection methods using Heaviside functions for sharper boundaries~\cite{Guest2004, Sigmund2007}. The converged SIMP solution yields a density field with predominantly binary values and localized intermediate densities at material interfaces. The element-wise nature of the resulting density field motivates the nodal reconstruction introduced in Section~\ref{sub:nodal-dense}.

\subsection{Level-set methods}\label{sub:levelset}

\noindent In contrast to density-based representations, level-set methods represent structural boundaries implicitly as the zero-level set of a continuous function, maintaining sharp boundary definition throughout optimization~\cite{VanDijk2013}.

\subsubsection{Implicit domain representation}

\noindent Consider a fixed computational domain $D \subset \mathbb{R}^3$. The structural domain $\Omega \subset D$ occupied by material is defined implicitly through a level-set function $\phi: D \rightarrow \mathbb{R}$ according to the convention~\cite{Allaire2004}:
\begin{equation}
    \begin{cases}
        \phi(\boldsymbol{x}) < 0 & \text{if } \boldsymbol{x} \in \Omega \text{ (solid)}\\
        \phi(\boldsymbol{x}) = 0 & \text{if } \boldsymbol{x} \in \partial\Omega \text{ (boundary)}\\
        \phi(\boldsymbol{x}) > 0 & \text{if } \boldsymbol{x} \in D \setminus \bar{\Omega} \text{ (void)}\,.
    \end{cases}
    \label{eqn:levelset_convention}
\end{equation}
The structural boundary $\partial\Omega$ corresponds to the zero-level set of $\phi$. This representation decouples the boundary representation from element connectivity~\cite{Osher1988}. In practice, $\phi$ is maintained as a signed distance function (SDF) of the current boundary $\partial\Omega$, defined as
\begin{equation}
    d_\Omega(\boldsymbol{x}) = 
    \begin{cases}
        -d(\boldsymbol{x}, \partial\Omega) & \text{if } \boldsymbol{x} \in \Omega\\
        0 & \text{if } \boldsymbol{x} \in \partial\Omega\\
        +d(\boldsymbol{x}, \partial\Omega) & \text{if } \boldsymbol{x} \in D \setminus \bar{\Omega}\,,
    \end{cases}
    \label{eqn:SDF}
\end{equation}
where $d(\boldsymbol{x}, \partial\Omega) = \min_{\boldsymbol{x}_p \in \partial\Omega} \|\boldsymbol{x} - \boldsymbol{x}_p\|$. The SDF satisfies the eikonal equation $|\boldsymbol{\nabla}\phi| = 1$ almost everywhere, which is a numerically favorable property for level-set evolution~\cite{Peng1999}. The procedure for maintaining this property during optimization is described in Section~\ref{subsub:reinitialization}.

\subsubsection{Ersatz material approximation}

\noindent The ersatz material approximation assigns material properties based on the level-set function~\cite{Wang2003, Allaire2004}. The approach employs a smooth characteristic function $I$ that interpolates between full material stiffness in solid regions and a small value $\varepsilon_0$ in void regions:
\begin{equation}
    I(\phi) = (1 - H_\eta(\phi)) + \varepsilon_0 H_\eta(\phi)\,,
    \label{eqn:ersatz}
\end{equation}
where $H_\eta$ is a regularized Heaviside function introducing a continuous transition over a band of half-width $\eta$ centered at the zero-level set~\cite{Osher2004}:
\begin{equation}
    H_\eta(\phi) = 
    \begin{cases}
        0 & \text{if } \phi < -\eta\\
        \displaystyle\frac{1}{2} + \frac{\phi}{2\eta} + \frac{1}{2\pi}\sin\!\left(\frac{\pi\phi}{\eta}\right) & \text{if } |\phi| \leq \eta\\
        1 & \text{if } \phi > \eta\,.
    \end{cases}
    \label{eqn:Heaviside}
\end{equation}
The bandwidth parameter $\eta$ is typically chosen proportional to element size~\cite{Wegert2025TopOpt}. Following the level-set convention~\eqref{eqn:levelset_convention}, solid regions ($\phi < 0$) yield $I \approx 1$ and void regions ($\phi > 0$) yield $I \approx \varepsilon_0$. The small value $\varepsilon_0$ prevents stiffness matrix singularity in void regions.

In the simplest implementation, the level-set function is evaluated at element centroids to obtain an element-wise constant material distribution~\cite{VanDijk2013}. Each element stiffness matrix becomes $\boldsymbol{k}_e(\phi) = I(\phi_e)\, \boldsymbol{k}_e^0$, with $\phi_e$ denoting the level-set value at the element centroid and $\boldsymbol{k}_e^0$ the element stiffness matrix for full material. The global equilibrium equation retains the standard form~\eqref{eqn:ls_equilibrium}:
\begin{equation}
    \mathbf{K}(\phi)\,\mathbf{U} = \mathbf{F}\,.
    \label{eqn:ls_equilibrium}
\end{equation}
Since the interpolation~\eqref{eqn:ersatz} is defined over the entire fixed domain~$D$, the finite element mesh remains unchanged throughout optimization, avoiding the need for remeshing or immersed boundary techniques~\cite{VanDijk2013}.

\subsubsection{Compliance minimization problem}

\noindent The compliance minimization problem for the level-set representation reads:
\begin{equation}
    \begin{aligned}
        \min_{\phi} \quad & J(\phi) = \mathbf{U}^\mathsf{T} \mathbf{K}(\phi)\, \mathbf{U} = \sum_{e=1}^{N} I(\phi_e)\, \boldsymbol{u}_e^\mathsf{T} \boldsymbol{k}_e^0\, \boldsymbol{u}_e \\
        \text{s.t.} \quad & \mathbf{K}(\phi)\,\mathbf{U} = \mathbf{F} \\
                          & V(\phi) \leq V^*\,,
    \end{aligned}
    \label{eqn:ls_compliance}
\end{equation}
where $N$ is the total number of elements and $V(\phi) = \sum_{e=1}^{N} V_e \bigl(1 - H_\eta(\phi_e)\bigr)$ is the material volume computed from the level-set function, with $V_e$ denoting the volume of element $e$. Unlike SIMP, which operates on element-wise density variables, the level-set formulation uses the nodal scalar field $\phi$ as the design variable~\cite{VanDijk2013}. Element-level stiffness is then obtained by evaluating $\phi$ at element centroids.

\subsubsection{Constraint handling}\label{subsub:constraint_handling}

\noindent The constrained problem~\eqref{eqn:ls_compliance} is solved using the augmented Lagrangian method~\cite{Nocedal2006}. Since the volume constraint is active at the optimum for compliance minimization, it is treated as an equality with target volume $V^* = V_f\, V_D$, where $V_f$ is the target volume fraction and $V_D$ is the volume of the computational domain. The normalized constraint residual $C(\phi) = \bigl(V(\phi) - V_f\, V_D\bigr) / V_D$ yields the augmented Lagrangian functional:
\begin{equation}
    \mathcal{L}(\phi, \lambda, \Lambda) = J(\phi) - \lambda\, C(\phi) + \frac{\Lambda}{2}\, C(\phi)^2\,,
    \label{eqn:augmented_Lagrangian}
\end{equation}
where $\lambda$ is the Lagrange multiplier and $\Lambda$ is the penalty parameter. Following~\cite{Nocedal2006}, the multiplier is updated at each iteration as:
\begin{equation}
    \lambda^{k+1} = \lambda^k - \Lambda^k\, C(\phi^k)\,.
    \label{eqn:lambda_update}
\end{equation}
The penalty parameter is increased periodically by a factor $\xi > 1$, subject to an upper bound $\Lambda_{\max}$~\cite{Wegert2025TopOpt}. When used with the Hilbertian extension--regularization (Section~\ref{subsub:hilbertian}), the objective and constraint sensitivities are first combined into the augmented Lagrangian sensitivity $g_{\mathcal{L}}$~\eqref{eqn:g_L}, then extended as a single field by solving~\eqref{eqn:hilbertian}. The augmented Lagrangian method can handle initial designs far from feasibility, but parameter tuning is problem-dependent~\cite{Allaire2021} and becomes increasingly difficult as the number of constraints grows~\cite{Wegert2023, VanDijk2013}.

An alternative is the Hilbertian projection method~\cite{Wegert2023}, which does not rely on the augmented Lagrangian formulation. Instead of combining sensitivities into a single field, the same extension problem is solved separately for the objective and for each constraint sensitivity. The method handles multiple and linearly dependent constraints through Gram--Schmidt orthogonalization, which naturally removes linearly dependent constraint sensitivities~\cite{Wegert2023}. In this work, only the volume equality constraint is present, and the general formulation reduces to a single pair of extended fields $g_\Omega \in H$ and $\mu_\Omega \in H$ for the objective and constraint sensitivity, respectively. The resulting velocity $v_\Omega$ is given by:
\begin{equation}
    v_\Omega = \sqrt{1 - \alpha^2}\;\frac{P_{\mathcal{C}^\perp}\, g_\Omega}{\|P_{\mathcal{C}^\perp}\, g_\Omega\|_H} + \alpha\, \frac{\mu_{\Omega}}{\|\mu_{\Omega}\|_H}\,,
    \label{eqn:hilbertian_projection}
\end{equation}
where $P_{\mathcal{C}^\perp}$ is the orthogonal projection onto the complement of the subspace $\mathcal{C} \subset H$ spanned by $\mu_\Omega$, and $\alpha$ is a scalar coefficient. The first term improves the objective without affecting the volume constraint to first order. The second term drives constraint satisfaction. The coefficient $\alpha$ is determined from the requirement $\langle \mu_\Omega,\, v_\Omega \rangle_H = \beta\, C$, where $\beta$ is a rate parameter for constraint correction, distinct from the Lagrange multiplier $\lambda$ in~\eqref{eqn:augmented_Lagrangian}. The coefficient is subject to the bounds $\alpha_{\min}^2 \leq \alpha^2 \leq 1$. The lower bound guarantees a minimum contribution of the constraint correction term. The upper bound ensures a real-valued objective improvement term. When the lower bound is active, $\beta$ is adjusted so that the requirement $\langle \mu_\Omega,\, v_\Omega \rangle_H = \beta\, C$ remains satisfied. For near-feasible initializations, the natural $\alpha^2$ can be orders of magnitude below $\alpha_{\min}^2$. The enforced minimum correction overshoots the nearly satisfied constraint, triggering alternating corrections in subsequent iterations. We address this oscillatory behavior by adaptively reducing the effective lower bound on $\alpha^2$ (Section~\ref{sub:setup}). The method in its current form handles only equality constraints~\cite{Wegert2023}.

\subsubsection{Shape derivative and domain relaxation}\label{subsub:shape_derivative}

\noindent Design updates in level-set optimization are driven by the shape derivative, which quantifies the sensitivity of $J(\Omega)$ to boundary perturbations~\cite{Allaire2004}. Compliance minimization requires no adjoint solve because the shape derivative depends only on the state $\boldsymbol{u}$~\cite{Allaire2004}. Considering the boundary $\partial\Omega$ with outward unit normal $\boldsymbol{n}$, the shape derivative for traction-free conditions under a perturbation field $\boldsymbol{\theta}$ (with normal component $v_n = \boldsymbol{\theta} \cdot \boldsymbol{n}$) is given by~\cite{Allaire2004}:
\begin{equation}
    J'(\Omega)(v_n) = -\int_{\partial\Omega} \boldsymbol{\sigma}(\boldsymbol{u}) : \boldsymbol{\varepsilon}(\boldsymbol{u})\, v_n\, \mathrm{d}s\,,
    \label{eqn:shape_derivative_integral}
\end{equation}
where $\boldsymbol{\sigma}(\boldsymbol{u}) : \boldsymbol{\varepsilon}(\boldsymbol{u})$ is the strain energy density. Note that the integrand is the continuous counterpart of the strain energy $\boldsymbol{u}_e^\mathsf{T} \boldsymbol{k}_e^0\, \boldsymbol{u}_e$ that, together with the penalization factor $p\rho_e^{p-1}$, drives the SIMP update~\eqref{eqn:sensitivity}. Thus, in both formulations, material is retained where the strain energy is highest.

Since the ersatz material framework does not explicitly represent the interface $\partial\Omega$ on the fixed mesh, the boundary integral~\eqref{eqn:shape_derivative_integral} is relaxed over~$D$ using an identity relating boundary and volume integrals~\cite{Osher2004, Wegert2025TopOpt}. To obtain a variational form compatible with the Hilbertian extension--regularization (Section~\ref{subsub:hilbertian}), the perturbation is parameterized as $\boldsymbol{\theta} = -w\,\boldsymbol{n}$ with a scalar test function $w \in H^1(D)$. The compliance sensitivity relaxed over $D$ then becomes:
\begin{equation}
    J'(\Omega)(-w\boldsymbol{n}) = \int_D \boldsymbol{\sigma}(\boldsymbol{u}) : \boldsymbol{\varepsilon}(\boldsymbol{u})\, w\, H'_\eta(\phi)\, |\boldsymbol{\nabla}\phi|\, \mathrm{d}\boldsymbol{x}\,,
    \label{eqn:dJ_relaxed}
\end{equation}
where $H'_\eta$ is the derivative of the smoothed Heaviside~\eqref{eqn:Heaviside}, which concentrates the integrand near $\partial\Omega$, and $|\boldsymbol{\nabla}\phi|$ accounts for the change of variables. This relaxation recovers the boundary integral in the limit $\eta \to 0$ when $\phi$ is a signed distance function ($|\boldsymbol{\nabla}\phi| = 1$), a property maintained through reinitialization (Section~\ref{subsub:reinitialization}). The volume constraint sensitivity takes an analogous form:
\begin{equation}
    C'(\Omega)(-w\boldsymbol{n}) = -\int_D \frac{1}{V_D}\, w\, H'_\eta(\phi)\, |\boldsymbol{\nabla}\phi|\, \mathrm{d}\boldsymbol{x}\,.
    \label{eqn:dC_relaxed}
\end{equation}
Combining the objective and constraint sensitivities according to the augmented Lagrangian~\eqref{eqn:augmented_Lagrangian}, the shape derivative of $\mathcal{L}$ with respect to a perturbation $-w\,\boldsymbol{n}$ reads:
\begin{equation}
    \mathcal{L}'(\Omega)(-w\boldsymbol{n}) = \int_D g_{\mathcal{L}}(\boldsymbol{x})\, w\, H'_\eta(\phi)\, |\boldsymbol{\nabla}\phi|\, \mathrm{d}\boldsymbol{x}\,,
    \label{eqn:dL_relaxed}
\end{equation}
where
\begin{equation}
    g_{\mathcal{L}}(\boldsymbol{x}) = \boldsymbol{\sigma}(\boldsymbol{u}) : \boldsymbol{\varepsilon}(\boldsymbol{u}) + \frac{\lambda - \Lambda\, C(\phi)}{V_D}
    \label{eqn:g_L}
\end{equation}
is the driving force for boundary motion. The first term promotes material retention in regions of high strain energy, while the second, modulated by the Lagrange multiplier and penalty, balances global material addition and removal to satisfy the volume constraint~\cite{Allaire2004}. Note that $g_{\mathcal{L}} > 0$ in structurally important regions, producing a positive normal velocity that advances the boundary into void ($\phi$ decreases), consistent with the convention $\phi < 0$ denoting solid. Standard finite element assembly of~\eqref{eqn:dL_relaxed} yields a nodal sensitivity vector $\mathbf{g}_{\mathcal{L}}$ that serves as the right-hand side for the velocity extension problem described next.

\subsubsection{Hilbertian extension-regularization}\label{subsub:hilbertian}

\noindent The sensitivity~\eqref{eqn:dL_relaxed} carries information concentrated near $\partial\Omega$ through the factor $H'_\eta(\phi)$. However, the Hamilton--Jacobi equation~\eqref{eqn:Hamilton-Jacobi} introduced in~Section~\ref{subsub:hamilton-jacobi} requires a smooth velocity field defined throughout $D$. The Hilbertian extension-regularization approach~\cite{Gournay2006} addresses both requirements by identifying the velocity $\bar{v}$ as the Riesz representative of $\mathcal{L}'$ in a weighted $H^1$ inner product. Specifically, one seeks $\bar{v} \in H^1(D)$ satisfying~\cite{Allaire2021, Wegert2025TopOpt}:
\begin{equation}
    \int_D \left( \ell^2 \boldsymbol{\nabla}\bar{v} \cdot \boldsymbol{\nabla} w + \bar{v}\, w \right) \mathrm{d}\boldsymbol{x} = \mathcal{L}'(\Omega)(-w\,\boldsymbol{n}) \quad \forall\, w \in H^1(D)\,,
    \label{eqn:hilbertian}
\end{equation}
where the right-hand side is given by~\eqref{eqn:dL_relaxed} and $\ell$ is a regularization length scale controlling the spatial smoothness of $\bar{v}$. Note that the smoothed velocity is denoted $\bar{v}$ to distinguish it from the raw normal component $\boldsymbol{\theta}\cdot\boldsymbol{n}$ used in the shape derivative. The $\ell^2$-regularized formulation~\eqref{eqn:hilbertian} extends the near-boundary sensitivity information into a globally smooth descent direction for $\mathcal{L}$~\cite{Allaire2021}. More generally, the right-hand side of~\eqref{eqn:hilbertian} can be replaced by any shape functional derivative. The Hilbertian projection method (Section~\ref{subsub:constraint_handling}) exploits this by solving the same system separately for $J'$ and each $C_i'$. The parameter $\ell$ is typically chosen proportional to the element size and the maximum boundary displacement per iteration. Since the design boundary must remain fixed where external loads are applied, homogeneous Dirichlet conditions $\bar{v} = 0$ are imposed on the traction (mechanical Neumann-type) boundary $\Gamma_N$~\cite{Allaire2004, Allaire2021}.

Finite element discretization of~\eqref{eqn:hilbertian} yields the linear system:
\begin{equation}
    \mathbf{A}_H\, \bar{\mathbf{v}} = \mathbf{g}_{\mathcal{L}}\,,
    \label{eqn:hilbertian_discrete}
\end{equation}
where $\bar{\mathbf{v}}$ is the nodal velocity vector and $\mathbf{A}_H$ is a symmetric positive-definite matrix arising from the $H^1$ inner product on the left-hand side of~\eqref{eqn:hilbertian}. The right-hand side $\mathbf{g}_{\mathcal{L}}$ is the nodal sensitivity vector obtained by finite element assembly of~\eqref{eqn:dL_relaxed}. Since $\mathbf{A}_H$ depends only on the mesh and the parameter $\ell$, we assemble and factor it once. The Hilbertian projection method (Section~\ref{subsub:constraint_handling}) reuses the same factorization, solving~\eqref{eqn:hilbertian_discrete} separately with the objective and constraint sensitivity vectors as right-hand sides. This step plays a role similar to sensitivity filtering in SIMP~\eqref{eqn:sensitivity_filter}: both regularize the raw sensitivity field to produce a well-conditioned design update. However, the Hilbertian extension simultaneously extends near-boundary information onto the entire domain~$D$, whereas the sensitivity filter smooths an already globally defined field~\cite{VanDijk2013, Allaire2021}.

\subsubsection{Hamilton--Jacobi evolution}\label{subsub:hamilton-jacobi}

\noindent Given the $\ell^2$-regularized velocity field $\bar{v}$ from the Hilbertian extension~\eqref{eqn:hilbertian}, the current level-set function $\phi^k$ is evolved according to the Hamilton--Jacobi equation~\cite{Osher1988, Sethian1999}:
\begin{equation}
    \frac{\partial \phi}{\partial t} + \bar{v}\,|\boldsymbol{\nabla}\phi| = 0\,.
    \label{eqn:Hamilton-Jacobi}
\end{equation}
The velocity field $\bar{v}$ is held fixed during the Hamilton--Jacobi evolution and is recomputed once per optimization iteration from the updated state~$\boldsymbol{u}$ via the shape derivative~\eqref{eqn:dL_relaxed} and the Hilbertian extension~\eqref{eqn:hilbertian}.

Consistent with the level-set convention~\eqref{eqn:levelset_convention} where $\phi < 0$ denotes solid, positive $\bar{v}$ advances the structural boundary into void regions while negative $\bar{v}$ retracts it~\cite{Osher2004}.

The equation is solved using forward Euler time integration with a first-order Godunov upwind spatial stencil on the Cartesian finite element mesh~\cite{Osher2004, Peng1999}. As the scheme is explicit, the time step $\Delta t$ must satisfy the Courant--Friedrichs--Lewy (CFL) condition:
\begin{equation}
    \Delta t \leq \frac{h}{\max|\bar{v}|}\,,
    \label{eqn:CFL}
\end{equation}
where $h$ denotes the element edge length. For the structured hexahedral meshes used in this work, $h$ is uniform across the domain. On general non-uniform meshes, the minimum edge length provides a conservative bound that ensures stability throughout. The evolution is performed for a prescribed number of steps $n_\mathrm{steps}$ with a CFL number $\gamma \in (0, 1)$ that controls the fraction of the stability limit used~\cite{Wegert2025TopOpt}. Combined with the CFL bound~\eqref{eqn:CFL}, the maximum boundary displacement per optimization iteration is bounded by $n_\mathrm{steps}\,\gamma\,h$. In contrast to the reinitialization equation (Section~\ref{subsub:reinitialization}), the Hamilton--Jacobi evolution is not iterated to pseudo-steady state but is performed for a fixed number of steps. The CFL number is adaptively reduced (by 25\%~\cite{Wegert2025TopOpt}) when successive values of $\mathcal{L}$ exhibit non-monotone behavior to suppress oscillations.

\subsubsection{Reinitialization}\label{subsub:reinitialization}

\noindent After evolution steps, the level-set function may deviate from the signed distance property~\eqref{eqn:SDF}. Strong variations of $|\boldsymbol{\nabla}\phi|$ near the zero-level contour degrade the numerical approximation of geometric quantities such as the boundary normal and curvature~\cite{Allaire2004, VanDijk2013}. Moreover, the relaxed shape derivative~\eqref{eqn:dJ_relaxed} was derived under the assumption $|\boldsymbol{\nabla}\phi| = 1$ and loses validity otherwise. Reinitialization restores the property $|\boldsymbol{\nabla}\phi| = 1$ almost everywhere by solving the following equation to steady state~\cite{Peng1999, Osher2004}:
\begin{equation}
    \frac{\partial \phi}{\partial t} + S(\phi_0)(|\boldsymbol{\nabla} \phi| - 1) = 0, \quad \phi(\boldsymbol{x}, 0) = \phi_0(\boldsymbol{x})\,,
    \label{eqn:reinitialization}
\end{equation}
where $\phi_0$ is the level-set function before reinitialization and $S(\phi_0)$ is its sign function~\cite{Peng1999}. In practice, the sign function is recomputed from the current $\phi$ at each pseudo-time step, which improves numerical stability near the interface~\cite{Peng1999, Wegert2025TopOpt}. Furthermore, the sign function $S$ is approximated by a smeared version~\cite{Peng1999}:
\begin{equation}
    S(\phi) = \frac{\phi}{\sqrt{\phi^2 + h^2\, |\boldsymbol{\nabla}\phi|^2}}\,,
    \label{eqn:smeared_sign}
\end{equation}
where $h$ is the minimum element edge length as in~\eqref{eqn:CFL}. The reinitialization equation~\eqref{eqn:reinitialization} is solved using the same upwind finite difference scheme as the Hamilton--Jacobi evolution. Reinitialization proceeds until the maximum absolute change in $\phi$ between successive pseudo-time steps falls below a prescribed tolerance. It is performed at each optimization iteration~\cite{Wegert2025TopOpt}.

\subsubsection{Initialization and topology limitations}\label{subsub:initialization}

\noindent Level-set optimization depends on the initial design. The Hamilton--Jacobi equation evolves existing boundaries but cannot create new ones in the material interior. Existing holes can merge or disappear, but new holes cannot nucleate. In two dimensions, the restriction is strict. In three dimensions, holes can connect and split through merging of zero-level surfaces~\cite{Allaire2004}, but results remain sensitive to initialization. This limitation is commonly addressed by providing an initial design with a suitable hole pattern~\cite{Allaire2004}. Topological derivative-based approaches~\cite{Allaire2005, Burger2004} enable hole nucleation but introduce user-defined parameters controlling the number, frequency, and placement of inserted holes~\cite{Barrera2020}. Improper settings can lead to excessive or insufficient nucleation, compromising both efficiency and final performance~\cite{Barrera2020}. Moreover, each insertion generates a discrete, non-differentiable step that may affect convergence~\cite{VanDijk2013}. Alternative formulations based on reaction-diffusion equations~\cite{Yamada2010} or density-based hole seeding~\cite{Barrera2020, Hoghoj2025} enable hole creation through natural evolution of the level-set function. Nevertheless, these methods still require careful tuning of continuation parameters, and a fully systematic, parameter-free procedure remains an open challenge~\cite{Barrera2020, VanDijk2013}.

\subsection{Motivation for sequential coupling}\label{sub:motivation}

\noindent The preceding sections reveal complementary characteristics that motivate the sequential combination of SIMP and level-set methods. Table~\ref{tab:comparison} summarizes the key properties of each approach.

\begin{table*}[h]
\centering
\caption{Comparison of SIMP and level-set method characteristics.}
\label{tab:comparison}
\begin{tabular}{lcc}
\toprule
\textbf{Property} & \textbf{SIMP} & \textbf{Level-set} \\
\midrule
Boundary definition & Diffuse & Sharp \\
Convergence & Fast & Slower~\cite{VanDijk2013} \\
Design update scope & Entire domain & Boundary vicinity~\cite{VanDijk2013} \\
Hole nucleation & Automatic & Requires mechanism \\
Initialization sensitivity & Low & High~\cite{Allaire2004} \\
Post-processing required & Yes & Minimal \\
\bottomrule
\end{tabular}
\end{table*}

SIMP excels at rapid topology exploration. The density formulation enables simultaneous material redistribution across the entire design domain, allowing both hole creation and elimination without dedicated nucleation mechanisms. However, the converged results require geometric interpretation because element-wise constant densities produce stair-stepped boundaries, and filtering introduces gradual density transitions~\cite{Subedi2020}.

Level-set methods maintain sharp boundary definition throughout optimization, producing interfaces suitable for manufacturing applications~\cite{VanDijk2013, Allaire2021}. However, convergence is typically slower because design updates are confined to the boundary vicinity, and the initialization limitations discussed in Section~\ref{subsub:initialization} constrain achievable topologies.

The sequential framework exploits this asymmetry. SIMP establishes a feasible, topologically informed configuration through simultaneous density updates across the entire domain. The SDF-based geometry extraction then transforms this density distribution into a well-conditioned implicit function~\cite{Peng1999} that initializes Hamilton--Jacobi evolution. This coupling addresses the initialization sensitivity of standalone level-set approaches, allowing the level-set stage to focus primarily on boundary refinement. Topological modifications can still occur, but proceed from a starting configuration with an already-developed topology.

\section{Methodology}\label{sect:methodology}

\noindent This section describes the proposed sequential topology optimization methodology. Section~\ref{sub:pipeline-overview} gives a pipeline overview and identifies the novel contributions, and Sections~\ref{sub:nodal-dense}--\ref{sub:Discretization} detail the algorithmic stages.

\subsection{Pipeline overview}\label{sub:pipeline-overview}

\begin{figure*}[h!]
  \centering
  \includegraphics[clip,width=0.95\textwidth]{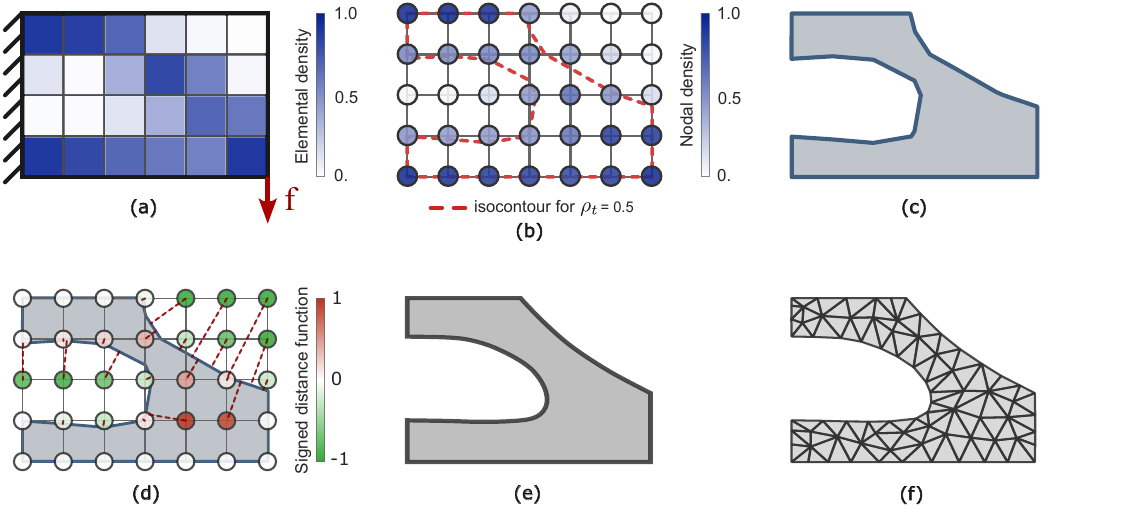}
  \caption{Illustration of the sequential topology optimization methodology on a 2D beam, used here for visual clarity (the methodology and all validation studies are performed in 3D). (a)~SIMP optimization result showing element-wise density distribution. (b)~Elemental densities mapped to nodes. (c)~Piecewise linear boundary extracted at the density threshold $\rho_t = 0.5$. (d)~SDF constructed on the finite element mesh nodes. (e)~Result of the level-set refinement stage, showing the sharpened boundary geometry. (f)~Final geometry discretization.}
  \label{fig:Sequential_pipeline}
\end{figure*}

\noindent This section presents the proposed sequential topology optimization methodology that combines density-based and level-set approaches. The main steps are illustrated in Figure~\ref{fig:Sequential_pipeline} and detailed in the following subsections. Steps 1 and 5 apply the standard SIMP and level-set formulations of Section~\ref{sect:background} without modification; the novel contributions of this work lie in the geometry transfer stages (steps 2--4), in which the SIMP density field is converted into an SDF suitable for level-set initialization. The methodology comprises:
\begin{enumerate}
  \item \textbf{SIMP topology optimization} (Figure~\ref{fig:Sequential_pipeline}a): The density-based optimization determines an initial material distribution within the design domain. The SIMP method is detailed in Section~\ref{sub:SIMP}.

  \item \textbf{Mapping elemental densities to nodes} (Figure~\ref{fig:Sequential_pipeline}b): The element-wise density values obtained from SIMP optimization are mapped to the nodes of the finite element mesh, yielding nodal density values required for geometry extraction. This mapping is detailed in Section~\ref{sub:nodal-dense}.

  \item \textbf{Iso-surface extraction} (Figure~\ref{fig:Sequential_pipeline}c): The material boundary is extracted as the iso-surface at density threshold $\rho_t = 0.5$, yielding a triangulated surface mesh. This threshold approximately preserves the prescribed volume fraction. Observed deviations are reported in Section~\ref{sect:num_tests}. The extraction process is described in Section~\ref{sub:geom_extraction}.

  \item \textbf{Signed distance function (SDF) construction} (Figure~\ref{fig:Sequential_pipeline}d): The extracted triangulated surface is converted into an SDF defined on the nodes of the original finite element mesh; see Section~\ref{sub:SDF} for details.

  \item \textbf{Level-set refinement} (Figure~\ref{fig:Sequential_pipeline}e): The constructed SDF initializes the level-set function. Hamilton--Jacobi evolution then refines the boundary and accommodates topological changes where required. The method is detailed in Section~\ref{sub:levelset} and its application to the sequential pipeline in Section~\ref{sub:levelset-opt}.

  \item \textbf{Discretization for engineering applications} (Figure~\ref{fig:Sequential_pipeline}f): The final optimized level-set function is converted into a computational mesh suitable for finite element analysis, CAD integration, or manufacturing processes. This step is covered in Section~\ref{sub:Discretization}.
\end{enumerate}

\subsection{Density field processing} \label{sub:nodal-dense}
\noindent The density field produced by SIMP (Section~\ref{sub:SIMP}) is element-wise constant. Its conversion to a continuous nodal field is the first step of the geometry transfer pipeline. Following~\cite{Jezek2026}, we employ a local linear least-squares fit that maps elemental densities onto mesh nodes. This fit reproduces linear density gradients exactly, whereas simple arithmetic averaging smears them. The difference is particularly pronounced on irregular meshes.

For each node, a linear polynomial approximates the local density field:
\begin{equation}
  \rho(\boldsymbol{x}) = [1 \quad \boldsymbol{x}^\mathsf{T}] \, \mathbf{a}\,,
  \label{eqn:lin_inter_lsq}
\end{equation}
where $\boldsymbol{x} \in \mathbb{R}^3$ is a spatial position and $\mathbf{a} = (a_0, a_1, a_2, a_3)^\mathsf{T} \in \mathbb{R}^4$ collects the constant and three linear coefficients. These coefficients are determined by solving the normal equations:
\begin{equation}
  \mathbf{X}^\mathsf{T} \mathbf{X} \, \mathbf{a} = \mathbf{X}^\mathsf{T} \boldsymbol{\rho}\,,
  \label{eqn:lin_inter_normaleq}
\end{equation}
where $\mathbf{X} \in \mathbb{R}^{m \times 4}$ stacks the rows $[1 \quad \boldsymbol{x}_e^\mathsf{T}]$ for the $m$ elements sharing the node and $\boldsymbol{\rho} \in \mathbb{R}^{m}$ collects their corresponding densities. The nodal density is then obtained by evaluating the fitted polynomial at the node location:
\begin{equation}
  \rho_n = [1 \quad \boldsymbol{x}_n^\mathsf{T}] \, \mathbf{a}\,.
  \label{eqn:nodal_density_eval}
\end{equation}

When $m \geq 4$ and the element centroids are not coplanar, the normal equations admit a unique solution. When $1 < m < 4$ (typically edge or corner nodes), $\mathbf{X}^\mathsf{T}\mathbf{X}$ is rank-deficient and we fall back to distance-weighted averaging with weights $w_i \propto 1/\|\boldsymbol{x}_n - \boldsymbol{x}_i\|$. For $m = 1$, the single neighbor's density is assigned directly. Since each nodal value depends only on its local element neighborhood, all nodes can be processed independently.

\subsection{Geometry extraction} \label{sub:geom_extraction}

\noindent The nodal density values established in Section~\ref{sub:nodal-dense} enable extraction of an explicit boundary representation. The material boundary is defined as the iso-surface at threshold $\rho_t = 0.5$.

We extract this iso-surface using the \textit{ParaView} visualization toolkit~\cite{Ahrens2005}, scripted through its Python interface (\textit{pvpython}), which implements cell-based iso-surface extraction applicable to both hexahedral and tetrahedral meshes. For hexahedral elements, the algorithm follows the Marching Cubes approach~\cite{Lorensen1987}. For tetrahedral elements, the Marching Tetrahedra variant~\cite{Doi1991} is employed. Both methods identify iso-surface intersections with element edges through linear interpolation of nodal values and generate a triangulated surface mesh. \textit{ParaView} was chosen for its established implementation on mixed hexahedral and tetrahedral meshes. The procedure could equivalently be realized through direct VTK bindings or a native Julia library such as \textit{Meshing.jl}. Our implementation supports Marching Tetrahedra, but the numerical validation in Section~\ref{sect:num_tests} is restricted to hexahedral meshes.

The threshold $\rho_t = 0.5$ approximately preserves the volume fraction prescribed during SIMP optimization. The small deviation arises from the discretization of the iso-surface extraction. Observed values are reported in Section~\ref{sect:num_tests}, and any discrepancy is corrected during the subsequent level-set optimization, which enforces the volume constraint explicitly.

The direct-SDF construction of~\cite{Jezek2026} requires remeshing onto a Cartesian background grid suitable for RBF-based smoothing. In the sequential setting this step is unnecessary, because the subsequent level-set optimization refines the boundary directly without remeshing. We therefore adopt the explicit-extraction route, which keeps the SDF on the original finite element mesh and delegates residual geometric refinement to the level-set stage.

\subsection{Boundary representation using signed distance function} \label{sub:SDF}

\noindent The extracted triangulated surface mesh provides an explicit boundary representation that must be converted into an SDF for subsequent level-set optimization. As established in Section~\ref{sub:levelset}, the SDF assigns to each point a value equal to the minimum Euclidean distance to the boundary, with a sign indicating material presence (see Eq.~\eqref{eqn:SDF}).

We compute the discrete SDF on the nodes of the original finite element mesh, maintaining consistency between the SIMP and level-set discretizations. The construction follows the approach of~\cite{Jezek2026} for distance computation, but introduces a simpler sign-assignment scheme enabled by evaluating the SDF on the original finite element mesh (Section~\ref{subsub:sign}). The sign convention is adapted to match~\cite{Allaire2004}. The construction comprises two independent operations: distance computation and sign assignment.

\subsubsection{Distance computation} \label{subsub:distance_function_formulation}
\noindent The distance from a mesh node $\boldsymbol{x}_n$ to the triangulated surface is computed by finding the closest point across all triangles. For each triangle $T_k$ with vertices $\{\boldsymbol{q}_1, \boldsymbol{q}_2, \boldsymbol{q}_3\}$, the closest point $\boldsymbol{x}^*_k$ minimizes the squared Euclidean distance subject to barycentric-coordinate constraints, yielding the constrained optimization problem:
\begin{equation}
    \begin{cases}
        \text{find} & \boldsymbol{\lambda} = \arg\min\limits_{\boldsymbol{\lambda}} \|\boldsymbol{x}_n - \boldsymbol{x}(\boldsymbol{\lambda})\|^2 \\[4pt]
        \text{s.t.} & \lambda_1 + \lambda_2 + \lambda_3 = 1 \\
        & \lambda_i \geq 0, \quad i = 1,2,3\,
    \end{cases}
    \label{eqn:closest_point}
\end{equation}
where $\boldsymbol{x}(\boldsymbol{\lambda}) = \sum_{i=1}^{3} \lambda_i \boldsymbol{q}_i$ represents a point on the triangle expressed in barycentric coordinates and the constraints ensure that the solution lies within or on the boundary of the triangle. Rather than solving~\eqref{eqn:closest_point} iteratively, we obtain $\boldsymbol{\lambda}^*_k$ in closed form by the case analysis detailed below. Denoting the optimal barycentric coordinates for triangle~$T_k$ as~$\boldsymbol{\lambda}^*_k$, the unsigned distance to the surface is computed as:
\begin{equation}
    d(\boldsymbol{x}_n) = \min_{k=1,\ldots,n_T} \|\boldsymbol{x}_n - \boldsymbol{x}(\boldsymbol{\lambda}^*_k)\|\,
    \label{eqn:distance_STL}
\end{equation}
where $n_T$ denotes the total number of triangles in the surface mesh.

The solution of~\eqref{eqn:closest_point} yields three projection types depending on the optimal barycentric coordinates. When all $\lambda_i > 0$, the unconstrained minimum lies in the triangle interior (orthogonal projection onto the triangle plane). If at least one constraint is active ($\lambda_i = 0$), the projection is clamped to the closest edge or vertex~\cite{Ericson2004}: an edge projection is computed by clamping the parametric coordinate along that edge to $[0, 1]$; a vertex projection is obtained when two constraints are active simultaneously. The clamping formulation remains well-defined even for narrow or near-degenerate triangles that may arise from iso-surface extraction: when the triangle plane is ill-conditioned, the projection naturally falls onto the shortest edge or nearest vertex.

Computational efficiency is achieved through spatial acceleration structures. A $k$-d tree constructed from triangle centroids~\cite{Bentley1975} provides candidate triangles within a search radius, implemented using the NearestNeighbors.jl library~\cite{NearestNeighbors}. AABB tests subsequently reject distant triangles before detailed projection computations. The per-node cost is $O(\log n_T + n_c)$, where $n_c$ is the number of candidate triangles examined, giving an overall SDF-construction cost of approximately $O(N_n \log n_T)$ for $N_n$ mesh nodes. As in the density-mapping step of Section~\ref{sub:nodal-dense}, each node is processed independently, enabling straightforward parallelization.

\subsubsection{Sign assignment}\label{subsub:sign}
\noindent The SDF is evaluated directly at nodes of the original finite element mesh, where processed nodal density values are available from Section~\ref{sub:nodal-dense}. Sign assignment therefore reduces to a simple thresholding operation. The signed distance function is obtained by combining the unsigned distance from Eq.~\eqref{eqn:distance_STL} with a sign determined by the nodal density:
\begin{equation}
    \phi(\boldsymbol{x}_n) =
    \begin{cases}
        -d(\boldsymbol{x}_n) & \text{if } \rho_n \geq \rho_t\,,\\[4pt]
        +d(\boldsymbol{x}_n) & \text{if } \rho_n < \rho_t\,,
    \end{cases}
    \label{eqn:signed_distance}
\end{equation}
where negative values indicate material presence, consistent with the level-set convention used throughout this work. Since the triangulated surface is extracted from the same nodal density field at the same threshold $\rho_t$, the density-based sign is consistent with the geometric inside/outside classification. For nodes near the boundary, where $\rho_n \approx \rho_t$, the unsigned distance approaches zero and the SDF value is insensitive to the assigned sign.

This approach contrasts with~\cite{Jezek2026}, where sign assignment required locating each grid node within the original mesh and evaluating the interpolated density field. Computing the SDF on the original mesh removes both operations, as nodal densities are directly accessible, which reduces implementation complexity and runtime overhead.

\subsubsection{Mesh-agnostic construction}\label{subsub:mesh_agnostic}

\noindent The geometry transfer described above operates on the original finite element mesh without uniform-spacing assumptions, and the underlying SDF construction was validated on hexahedral and tetrahedral discretizations in~\cite{Jezek2026}. The structured-mesh requirement of the present validation studies originates from the level-set stage (Section~\ref{sub:levelset-opt}). Extension to unfitted level-set formulations would exploit this property and remove the structured-mesh constraint. This direction is identified as future work in Section~\ref{sect:conclusion}.

\subsection{Level-set refinement}\label{sub:levelset-opt}

\noindent The SDF constructed in Section~\ref{sub:SDF} initializes the level-set function~$\phi$ for the shape-derivative--driven optimization of Section~\ref{sub:levelset}. The optimization is implemented using the \textit{GridapTopOpt.jl} package~\cite{Wegert2025TopOpt}, built on the \textit{Gridap.jl} finite element ecosystem~\cite{Badia2020, Verdugo2022}. The choice of constraint handler, an associated step-size modification, and other problem-specific parameters are reported in Section~\ref{sub:setup}.

\subsection{Final geometry discretization} \label{sub:Discretization}

\noindent The optimized level-set function provides an implicit geometry representation through its zero-level set. Conversion to explicit discrete meshes is required for finite element analysis, CAD integration, and manufacturing.

Surface discretization employs the Marching Cubes algorithm~\cite{Lorensen1987}, as used earlier in the SIMP geometry extraction step (Section~\ref{sub:geom_extraction}), now applied to the optimized SDF to extract a triangulated surface mesh (STL format). For volumetric discretization, we use \textit{TetGen}~\cite{Si2015} to generate a constrained Delaunay tetrahedralization that conforms to this boundary surface. Linear edge interpolation during surface extraction positions vertices on the zero-level set, enabling accurate representation of thin members and sharp features. No post-extraction surface smoothing is applied. The SDF is already regularized by the eikonal reinitialization (Section~\ref{subsub:reinitialization}), which produces sufficiently regular iso-surfaces for direct tetrahedralization.

This two-stage approach prioritizes geometric fidelity over guaranteed element quality bounds. Unlike the Isosurface Stuffing algorithm~\cite{Labelle2007}, which provides dihedral angle guarantees but constrains vertices to a background lattice and therefore smooths thin members, our approach places vertices precisely on the zero-level set, preserving features at sub-element resolution. The resulting mesh quality proved sufficient for the linear-elasticity validation analyses reported in Section~\ref{sect:num_tests}. \textit{TetGen}'s quality flags were set to their default values.

Beyond serving as the pipeline's output, we apply this same extraction--tetrahedralization procedure to compute the compliance values reported in Section~\ref{sect:num_tests}, ensuring that all compliance comparisons across stages and initialization strategies are performed on geometrically consistent discretizations (see Section~\ref{sub:setup} for details).

\section{Numerical tests and validation}\label{sect:num_tests}

\noindent The sequential methodology is assessed through two three-dimensional benchmark problems: a cantilever beam with corner supports and a Messerschmitt-B\"{o}lkow-Blohm (MBB) beam. These cases, with distinct topological characteristics (corner-supported versus symmetry-constrained), examine how SIMP-based initialization affects subsequent level-set refinement compared to the standard porous initialization. The scope is deliberately limited: a systematic parameter study across a broader benchmark library is outside the scope of this methodological paper and is identified as a direction for future work (Section~\ref{sect:conclusion}).

\subsection{Computational setup}\label{sub:setup}

\noindent To enable direct comparison, both problems share the same computational domain and parameters, differing only in boundary conditions as described in the following subsections. The domain spans $2.0 \times 1.0 \times 1.0$ in dimensionless units, paired with the unit load and material constants below to form a self-consistent validation setup. It is discretized using a structured hexahedral mesh with $40 \times 20 \times 20$ elements, a resolution that keeps the methodology comparison tractable on a single core while exercising all components of the pipeline (SIMP, SDF extraction, level-set refinement, and final discretization). The material has Young's modulus $E_0 = 1$~MPa and Poisson's ratio $\nu = 0.3$. All computations were performed on a single core of an Apple M3 Pro CPU. The level-set and SIMP stages are implemented in Julia 1.9~\cite{Bezanson2017} using \textit{GridapTopOpt.jl} 0.4.1, built on the \textit{Gridap.jl} 0.19 PDE solver framework. Geometry extraction and meshing use \textit{ParaView} 6.1.0 and \textit{TetGen} 1.6.

SIMP optimization is initialized with a uniform density field $\rho_e = V_f = 0.4$ and employs penalization exponent $p = 3$, sensitivity filter with radius $R = 2h$ where $h$ is the element edge length, and target volume fraction $V_f = 0.4$. The density field is updated using the Optimality Criteria scheme. Convergence is monitored through the maximum density change over all elements $\Delta\rho_{\max}$, with threshold values ranging from $0.5\%$ to $8\%$ to examine how SIMP convergence affects subsequent level-set refinement.
 
Two initialization strategies are compared for the level-set stage. The sequential approach (SIMP$\rightarrow$LS) extracts the SIMP density field as an SDF using the methodology described in Section~\ref{sub:SDF}. The baseline approach initializes from the default periodic porous design. The two strategies also differ in constraint handling. The sequential strategy employs Hilbertian projection~\cite{Wegert2023} with the step-size lower bound $\alpha_{\min}$ reduced adaptively as
\begin{equation}
    \alpha_{\min}^{\mathrm{eff}} = \alpha_{\min} \cdot \min\!\left(1,\; \frac{|C|}{\tau}\right),
    \label{eqn:alpha_min_decay}
\end{equation}
where $C$ is the volume constraint residual and $\tau = 0.01$ is a decay tolerance. The default lower bound $\alpha_{\min}^2 = 0.1$ follows the \textit{GridapTopOpt.jl} convention. Equation~\eqref{eqn:alpha_min_decay} reduces the step size as the design approaches feasibility, suppressing the oscillatory behavior described in Section~\ref{subsub:constraint_handling}. The multiplicative form was selected empirically. Alternative decay laws were not investigated. The porous initialization uses the augmented Lagrangian method. This pairing reflects the observed convergence behavior under default solver parameters. Hilbertian projection converges efficiently from near-feasible SIMP-derived initializations but did not converge from the porous initial design. The augmented Lagrangian method handles infeasible initializations more reliably but converged slowly from near-feasible starting points. A factorial comparison across both solvers and initializations was not pursued. The adaptive $\alpha_{\min}$ reduction in Eq.~\eqref{eqn:alpha_min_decay} is specific to the Hilbertian projection method and is not applied to the augmented Lagrangian baseline. The speedup reported in subsequent sections therefore reflects the full sequential framework (SIMP initialization, Hilbertian projection, and adaptive step-size modification), not initialization alone.

Both strategies use the default parameter settings of \textit{GridapTopOpt.jl}. For the sequential strategy, the convergence tolerance on the relative change of the objective $J$ between consecutive iterations is tightened by a factor of five relative to the default $J_{\mathrm{tol}} = 0.2\, h_{\max}$ (i.e.\ to $J_{\mathrm{tol}}/5 = 0.002$ at the present resolution). The default value yielded less converged results than the augmented Lagrangian baseline. The per-iteration cost differs between the two formulations. Hilbertian projection solves separate extension--regularization problems for the objective and constraint sensitivities, whereas the augmented Lagrangian combines them into a single extension problem. The additional linear system solve makes each Hilbertian projection iteration more expensive. The speedup values reported in subsequent sections are based on total wall-clock time and thus reflect this difference.
 
For consistent comparison across optimization stages, compliance values reported in the following sections are computed from geometries extracted using Marching Cubes and discretized with \textit{TetGen} (Section~\ref{sub:Discretization}). This extraction-based evaluation is necessary because the SIMP compliance includes contributions from penalized intermediate-density elements that are absent in the extracted geometry, making direct comparison misleading~\cite{Huang2021}. Although the level set provides a sharp geometric boundary, the ersatz-material approach approximates the stiffness of cut elements (those intersected by the zero level set) by weighting it with their solid volume fraction~\cite{VanDijk2013}, producing intermediate-density contributions analogous to those in SIMP but confined to a narrow band around the interface~\cite{Schmidt2024}. Applying the same extraction pipeline therefore enables direct comparison across all stages and initialization strategies (SIMP, LS with porous initialization, LS with SIMP initialization) under the same discretization procedure.

\subsection{Cantilever beam with corner supports}\label{sub:cantilever}

\begin{figure}[h!]
    \centering
    \includegraphics[width=0.3\textwidth]{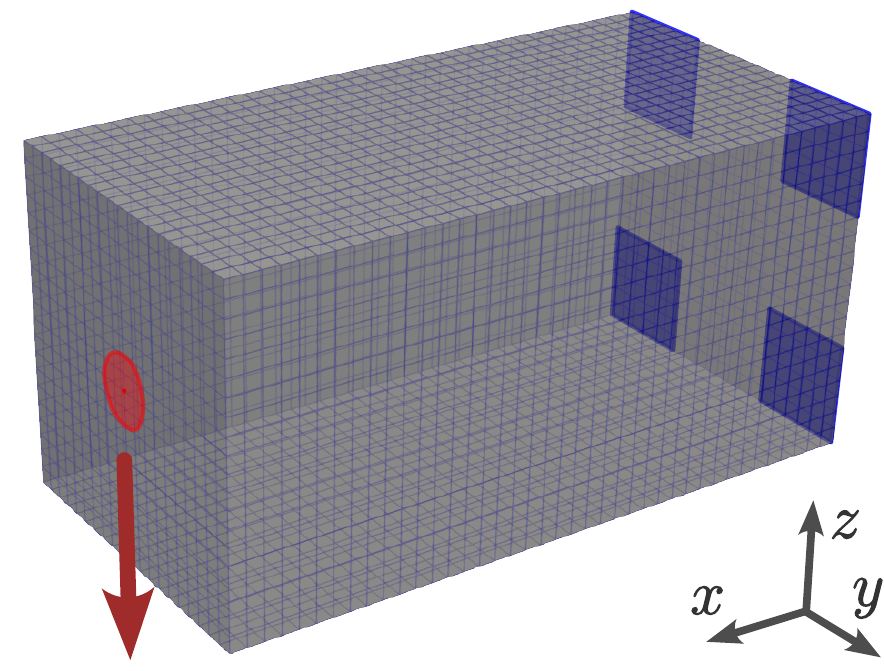}
    \caption{Cantilever beam problem setup: fixed supports at four corner 
regions (blue) and load applied to a circular region at the 
opposite face (red).}
    \label{fig:4Legs_geometry}
\end{figure}

\noindent The first test case considers a cantilever beam fixed at four corner regions on the $x = 0$ face, each occupying a $0.3 \times 0.3$ square. A load $\boldsymbol{f} = (0, 0, -1)$~N is applied to a circular region of radius $0.1$ centered at $(2.0, 0.5, 0.5)$ on the opposite face. Figure~\ref{fig:4Legs_geometry} illustrates the problem setup.

Tables~\ref{tab:detailed_comparison-4Legs} and~\ref{tab:seq_method-4Legs} present the optimization results. In the SIMP stage, tighter convergence criteria require more iterations but produce lower compliance in extracted geometries. The level-set stage requires substantially fewer iterations for all sequential initializations than for the porous baseline (Table~\ref{tab:detailed_comparison-4Legs}). The level-set iteration counts do not decrease monotonically with tighter SIMP convergence, as different initializations lead the level-set stage to nearby but distinct local minima. All final level-set compliance values, including the porous baseline, fall within a narrow range (spread below 2\%). The level-set stage therefore narrows the compliance differences established by the SIMP initialization. The advantage of the sequential framework lies in wall-clock time rather than in final compliance quality. All sequential configurations achieve speedup over the porous baseline, with the fastest ($\Delta\rho_{\max} = 8\%$) reaching $4.6\times$. This maximum speedup is obtained at the loosest SIMP convergence and at the highest final level-set compliance among the tested cases (67.59~J versus 66.35~J at the tightest criterion). The intermediate range $\Delta\rho_{\max} \in [1\%, 4\%]$ represents a practical trade-off, achieving $2.4$--$2.9\times$ speedup at compliance values comparable to the porous baseline. Consistent with the per-iteration cost discussion in Section~\ref{sub:setup}, Hilbertian projection iterations are roughly $1.4\times$ more expensive than augmented Lagrangian iterations in these tests (approximately $14$~s/iter versus $10$~s/iter), a difference already reflected in the reported wall-clock speedups. The geometry extraction and SDF construction add approximately $3.8$~s, negligible relative to the optimization stages.

\begin{figure*}[!t]
    \centering
    
    \fbox{\begin{minipage}[b]{0.47\textwidth}
        \centering
        \includegraphics[width=0.48\textwidth]{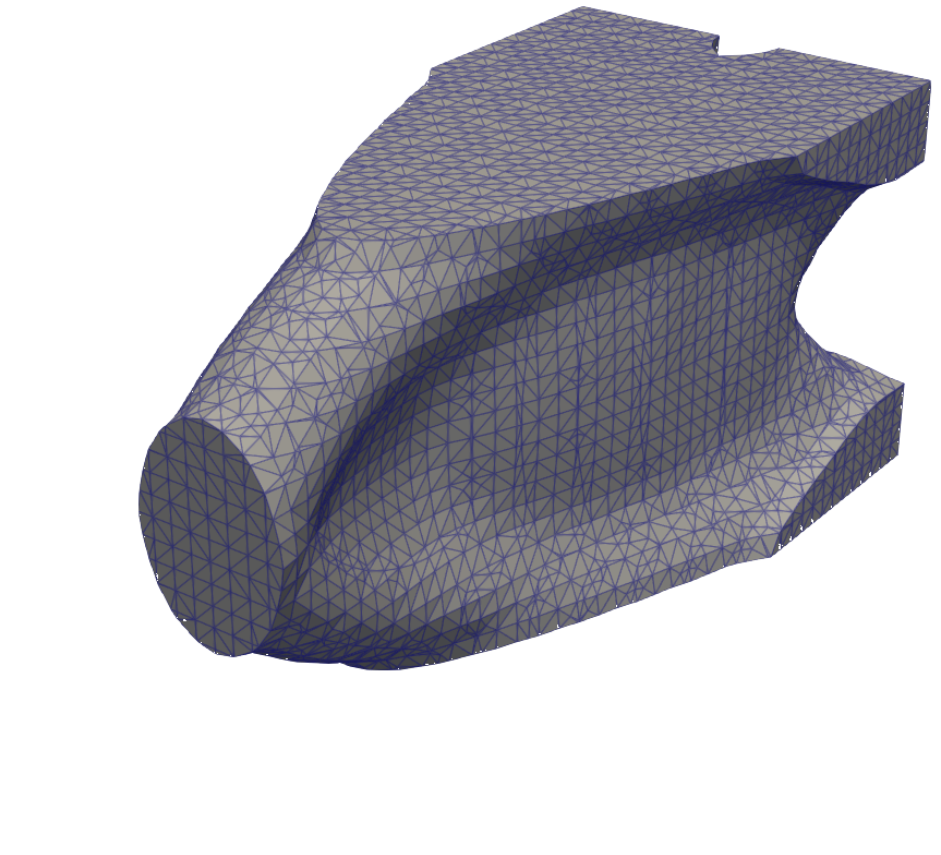}
        \hfill
        \includegraphics[width=0.48\textwidth]{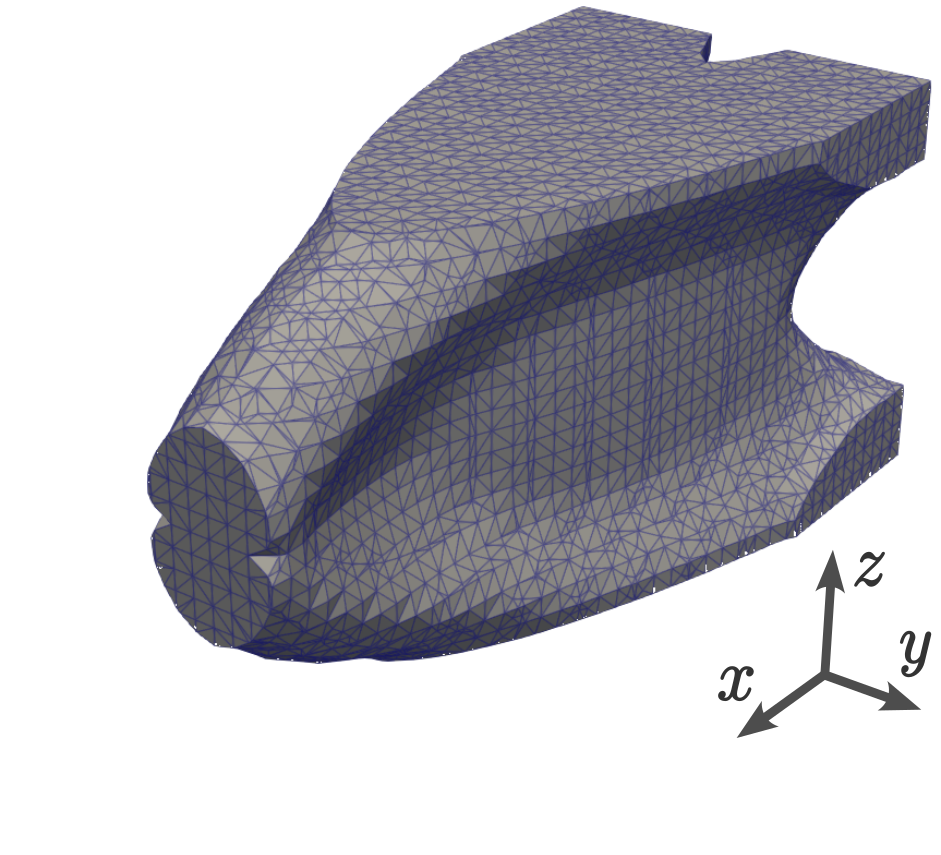}
        \\[0.1cm]
        \small\textbf{(a) $\Delta\rho_{\max} = 0.5\%$}
    \end{minipage}}
    \hfill
    \fbox{\begin{minipage}[b]{0.47\textwidth}
        \centering
        \includegraphics[width=0.48\textwidth]{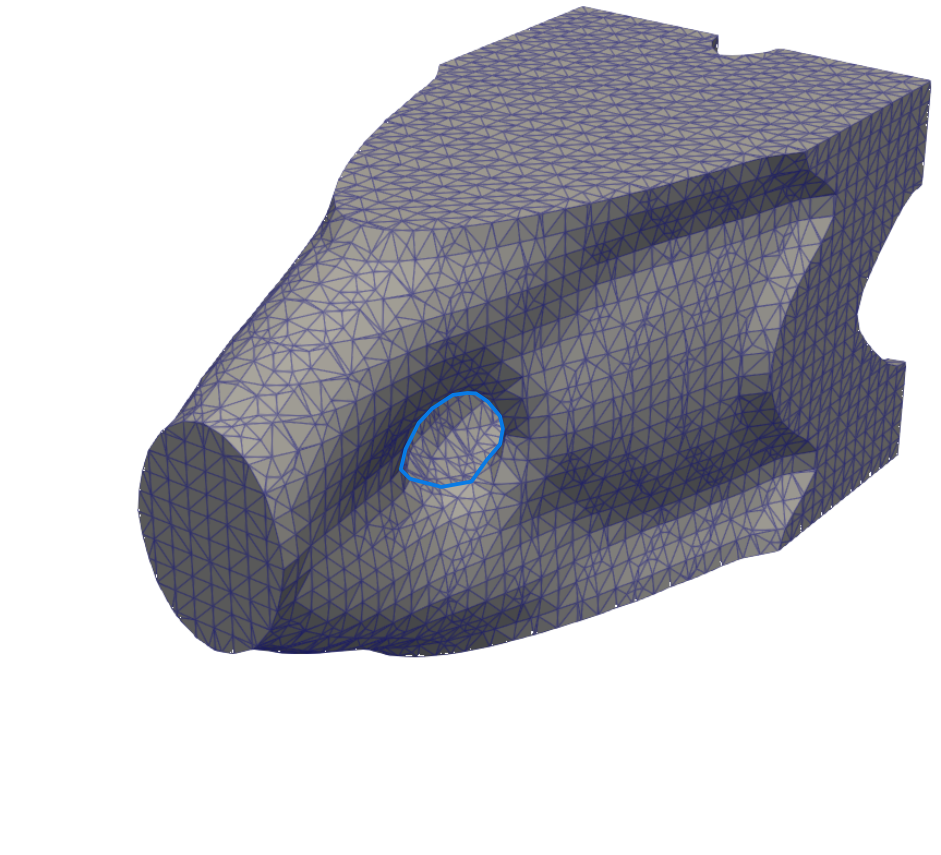}
        \hfill
        \includegraphics[width=0.48\textwidth]{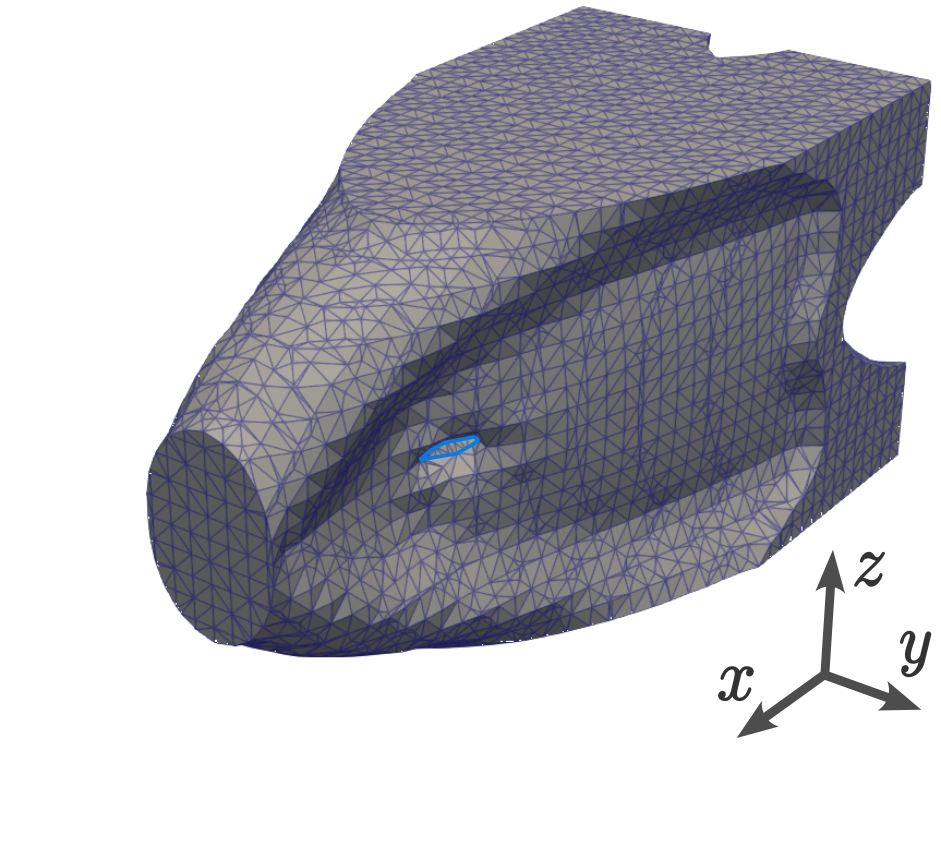}
        \\[0.1cm]
        \small\textbf{(b) $\Delta\rho_{\max} = 1\%$}
    \end{minipage}}

    \vspace{0.3cm}
    \fbox{\begin{minipage}[b]{0.47\textwidth}
        \centering
        \includegraphics[width=0.48\textwidth]{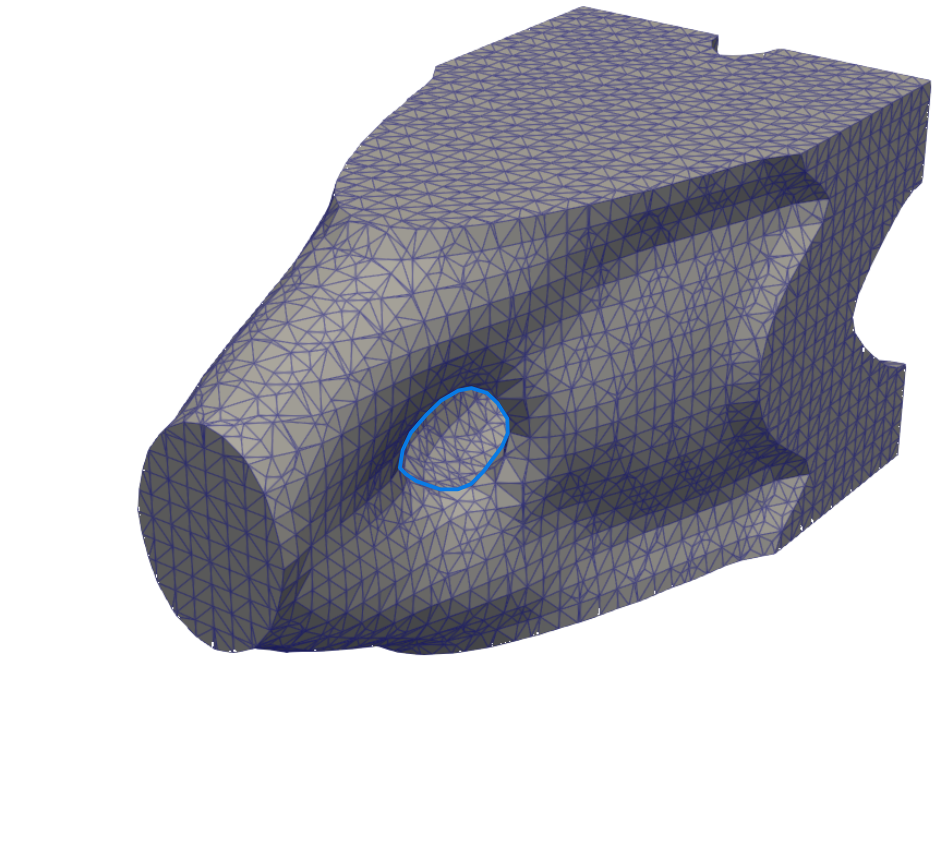}
        \hfill
        \includegraphics[width=0.48\textwidth]{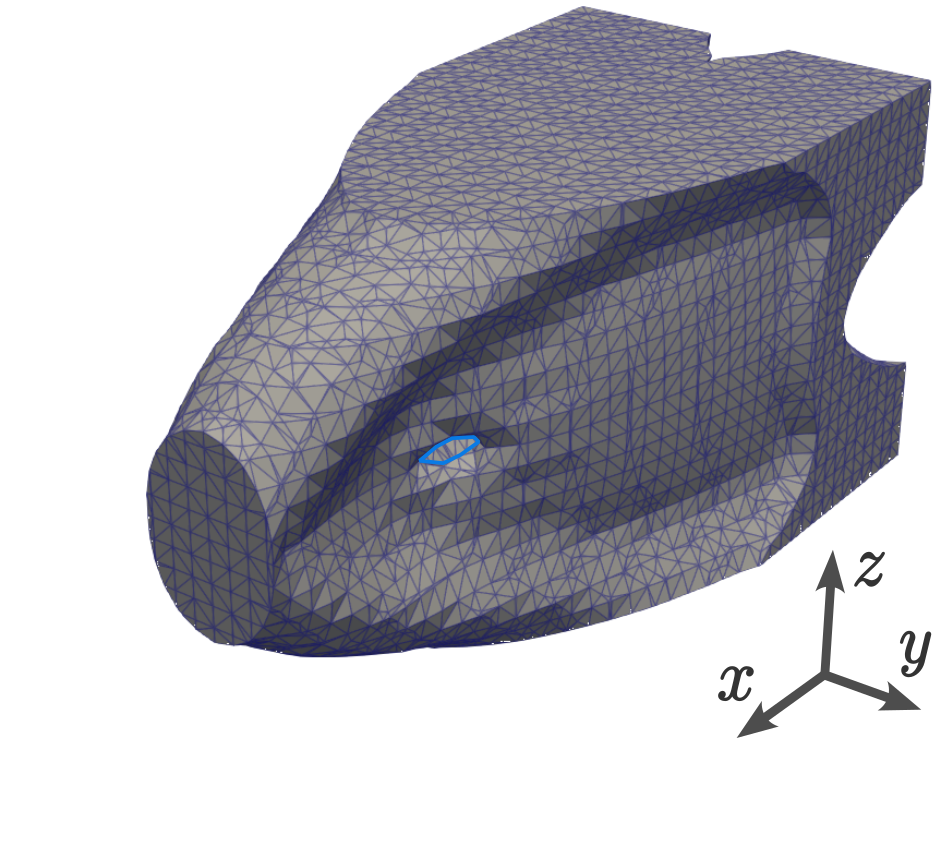}
        \\[0.1cm]
        \small\textbf{(c) $\Delta\rho_{\max} = 2\%$}
    \end{minipage}}
    \hfill
    \fbox{\begin{minipage}[b]{0.47\textwidth}
        \centering
        \includegraphics[width=0.48\textwidth]{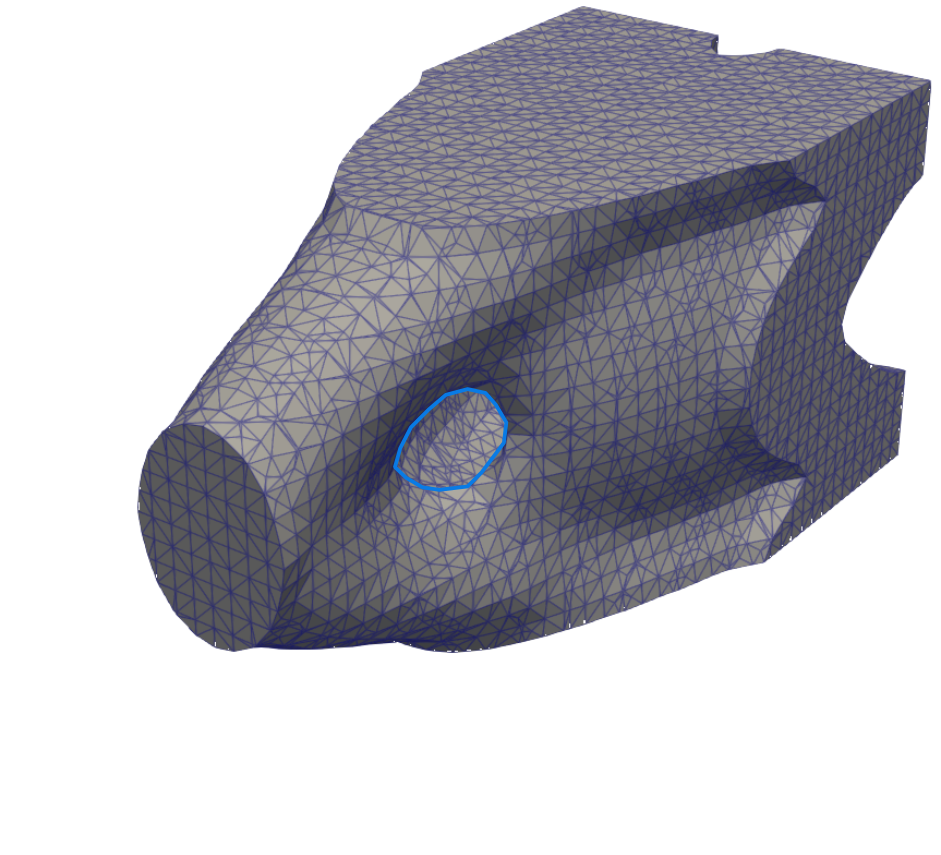}
        \hfill
        \includegraphics[width=0.48\textwidth]{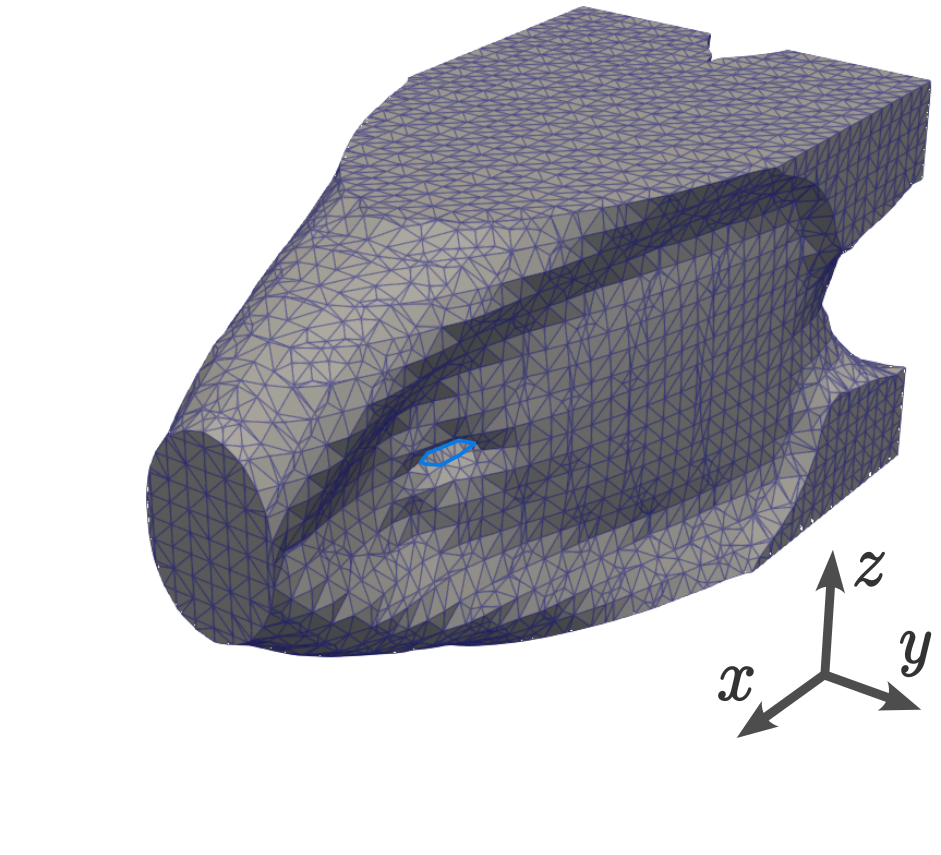}
        \\[0.1cm]
        \small\textbf{(d) $\Delta\rho_{\max} = 4\%$}
    \end{minipage}}
    \vspace{0.3cm}

    \fbox{\begin{minipage}[b]{0.47\textwidth}
        \centering
        \includegraphics[width=0.48\textwidth]{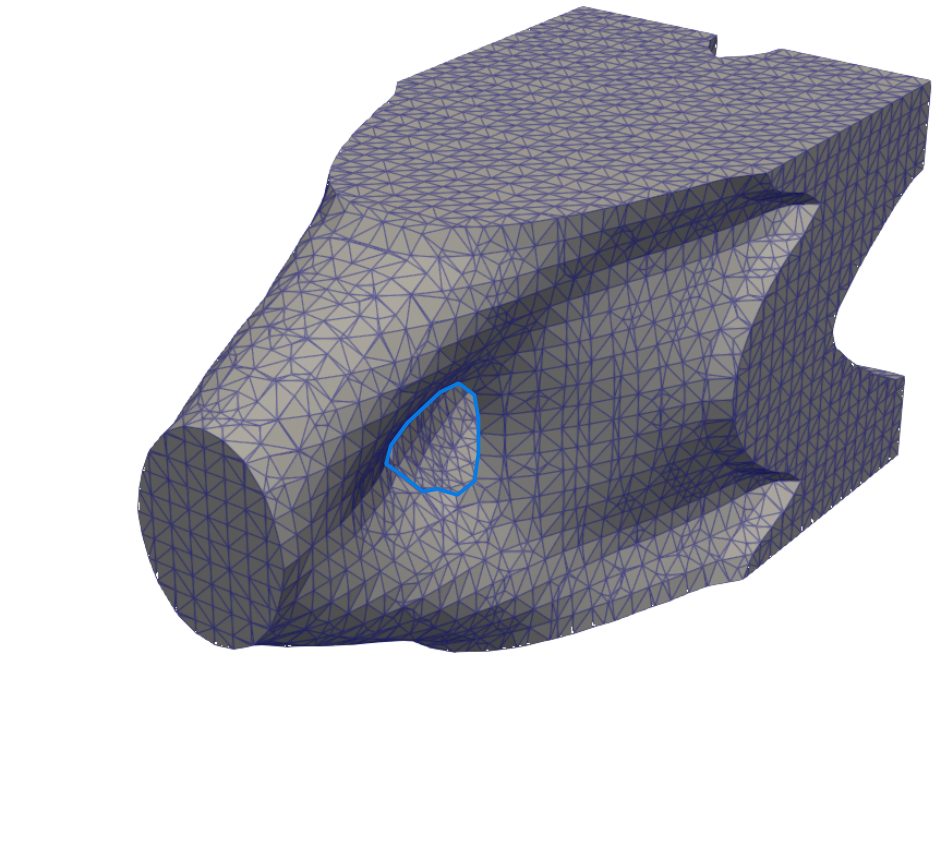}
        \hfill
        \includegraphics[width=0.48\textwidth]{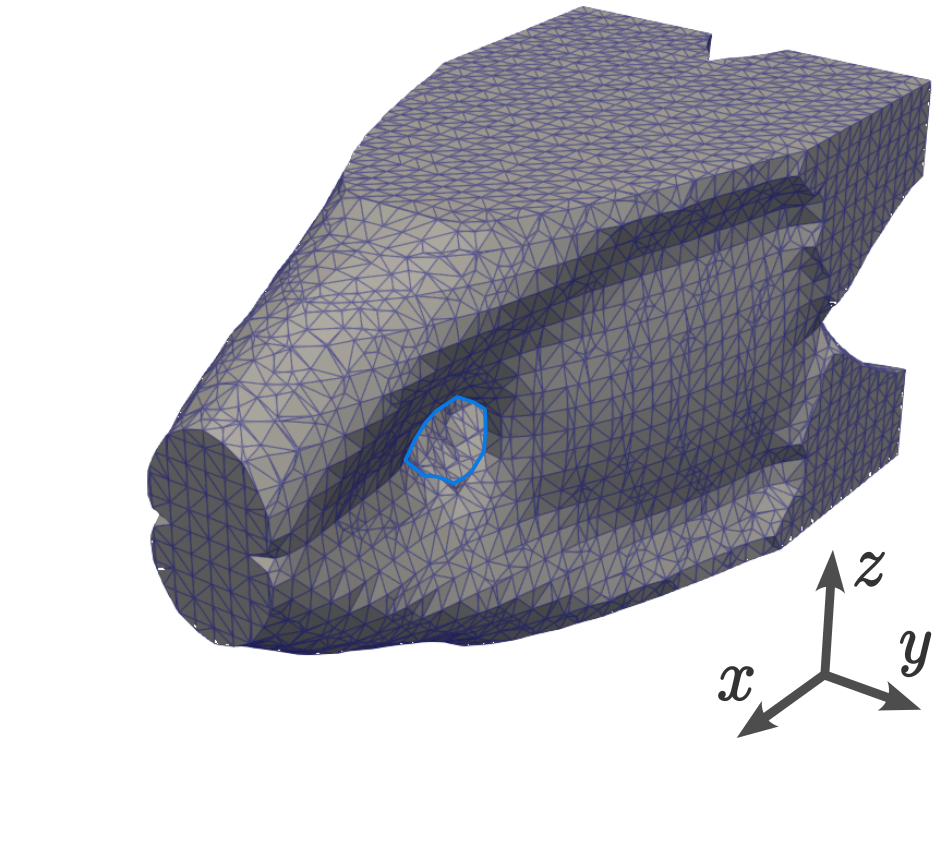}
        \\[0.1cm]
        \small\textbf{(e) $\Delta\rho_{\max} = 8\%$}
    \end{minipage}}
    \hfill
    \fbox{\begin{minipage}[b]{0.47\textwidth}
        \centering
        \includegraphics[width=0.48\textwidth]{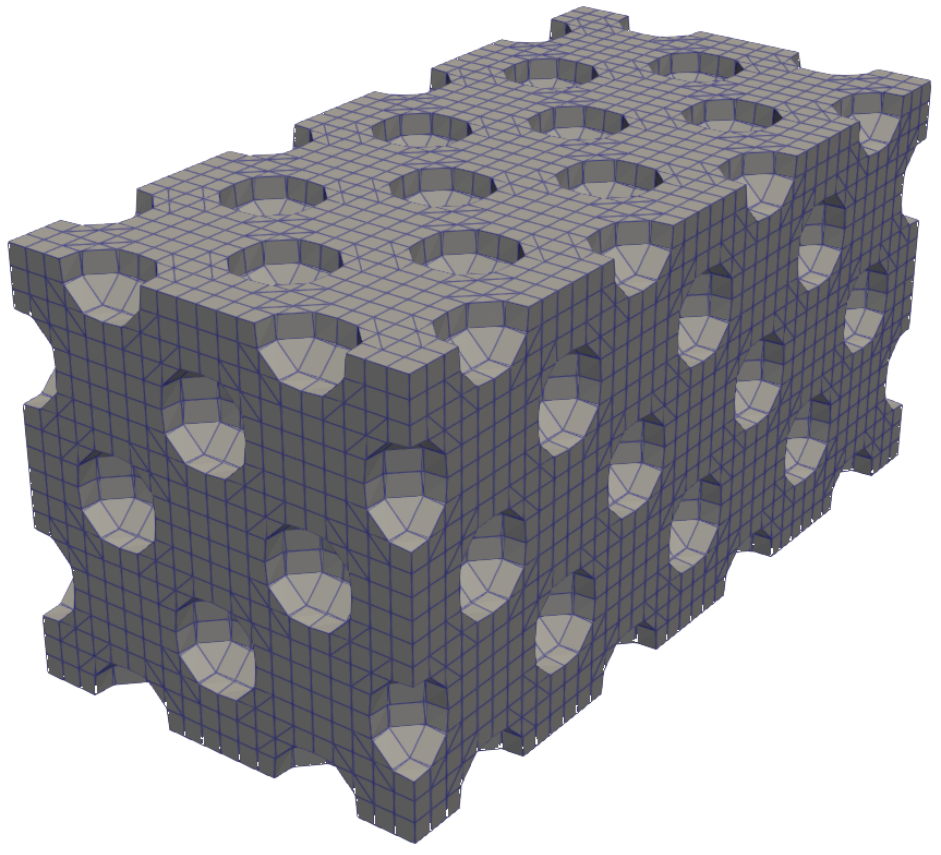}
        \hfill
        \includegraphics[width=0.48\textwidth]{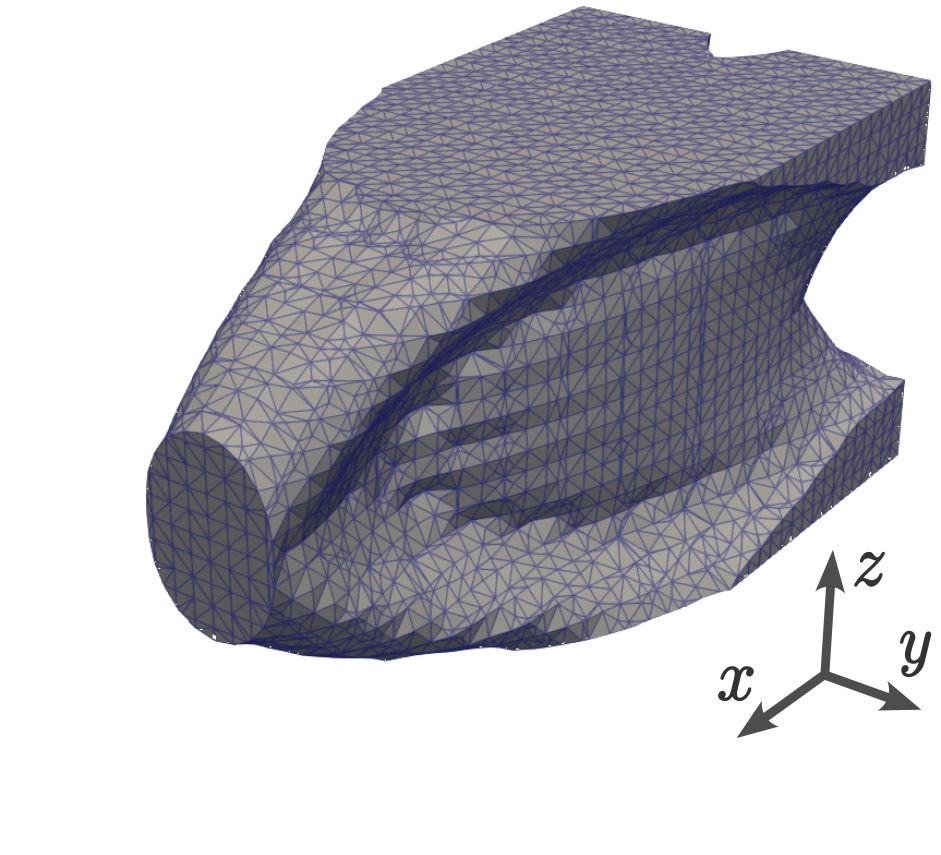}
        \\[0.1cm]
        \small\textbf{(f) Porous initialization}
    \end{minipage}}
    
    \vspace{0.2cm}
    \footnotesize Each box: SIMP extracted geometry (left) $\rightarrow$ level-set refined result (right).
    
    \caption{Cantilever beam optimization results. Each pair shows the extracted SIMP geometry (left) and the level-set refined result (right). Cases (a)--(e): sequential SIMP$\rightarrow$LS with varying convergence tolerance $\Delta\rho_{\max}$. Case (f): baseline from uniform porous initialization. Holes are marked to highlight topological differences between cases.}
    \label{fig:cantilever_results}
\end{figure*}

Figure~\ref{fig:cantilever_results} presents the resulting geometries. Each pair displays the extracted SIMP geometry (left) and the level-set refined result (right). Tighter SIMP convergence produces extracted geometries that progressively approach the level-set optimized shape. The level-set stage refines all initializations toward geometries resembling the most converged SIMP result ($\Delta\rho_{\max} = 0.5\%$). For this tightest convergence criterion, the level-set stage introduces only minor geometric modifications, indicating that well-converged SIMP optimization already identifies a geometry close to the level-set optimum for this problem. All cases produce comparable final topologies.

\begin{figure*}[!t]
    \centering
    \begin{subfigure}[b]{0.48\textwidth}
        \centering
        \includegraphics[width=\textwidth]{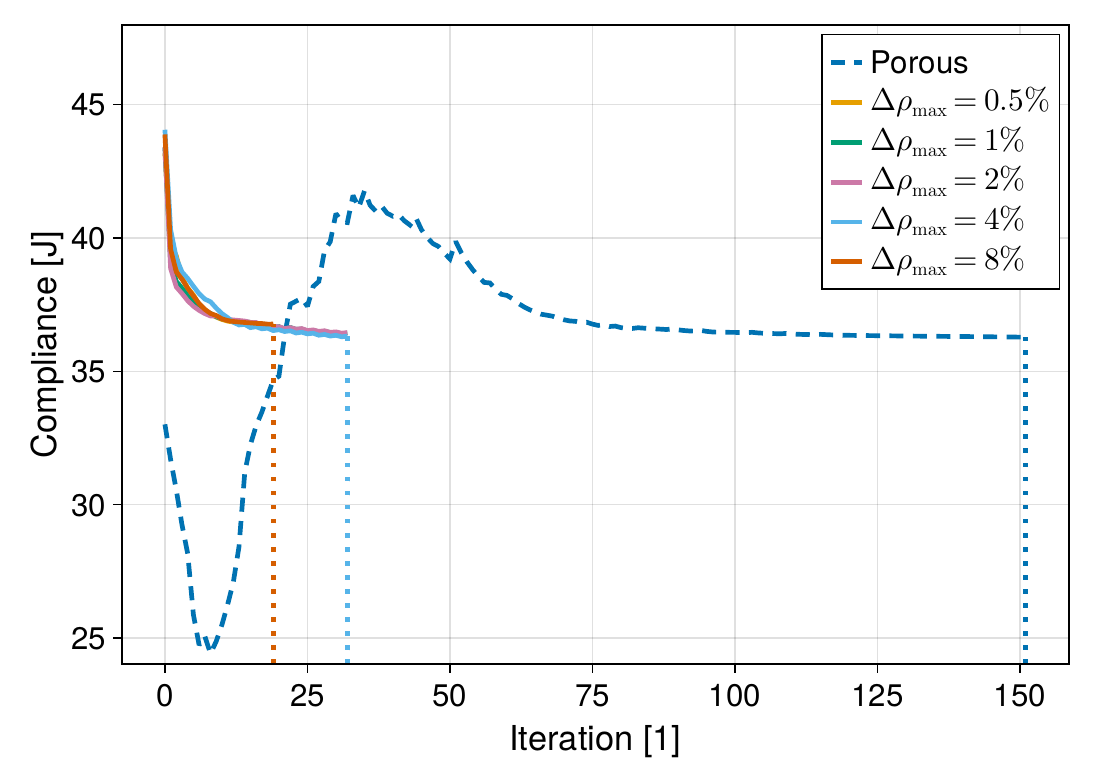}
        \caption{Compliance}
        \label{fig:4Legs_conv_J}
    \end{subfigure}
    \hfill
    \begin{subfigure}[b]{0.48\textwidth}
        \centering
        \includegraphics[width=\textwidth]{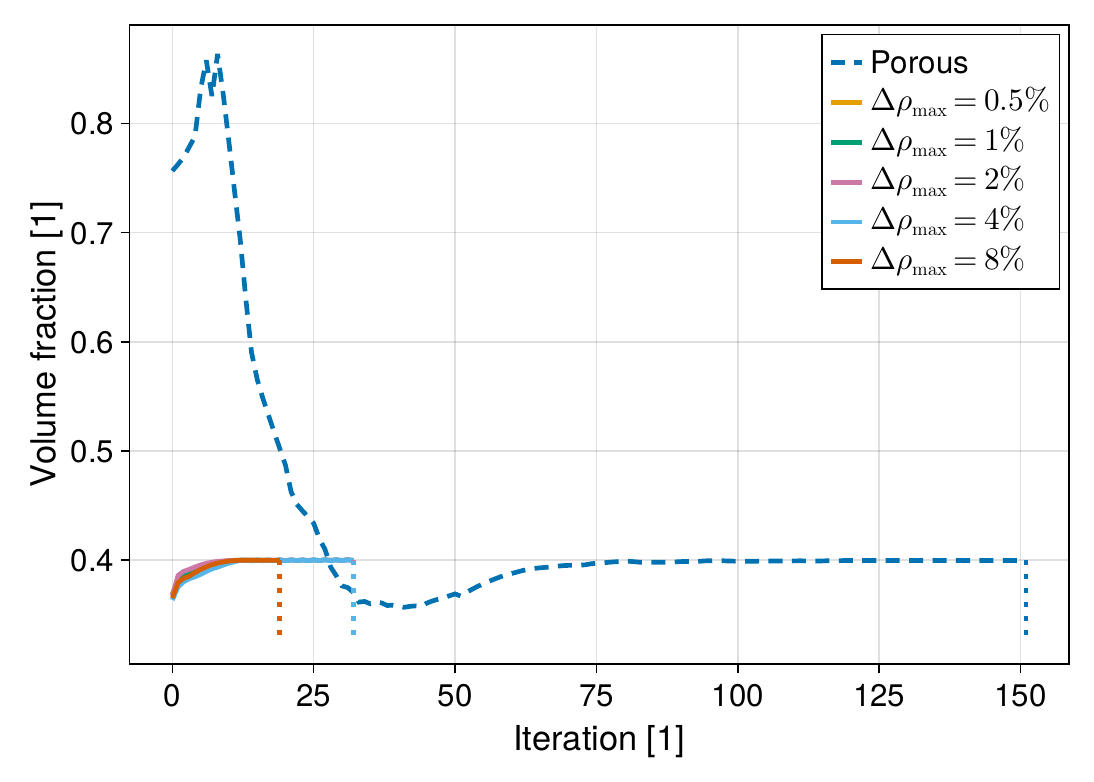}
        \caption{Volume fraction}
        \label{fig:4Legs_conv_Vol}
    \end{subfigure}
    \caption{Cantilever beam level-set convergence histories: (a)~Compliance and (b)~volume fraction of extracted geometries. Sequential initializations (SIMP$\rightarrow$LS) with varying $\Delta\rho_{\max}$ are compared against the porous initialization baseline (dashed line).}
    \label{fig:4Legs_convergence}
\end{figure*}

Figure~\ref{fig:4Legs_convergence} shows the level-set convergence histories. Sequential initializations converge faster and require fewer iterations than the porous initialization baseline (Figure~\ref{fig:4Legs_conv_J}). The volume fraction plot (Figure~\ref{fig:4Legs_conv_Vol}) shows that sequential initializations start close to the prescribed volume fraction, with approximately $3\%$ deviation caused by the reinitialization step before optimization. The porous initialization starts far from the target volume and requires more iterations to converge.

\begin{table*}[!t]
\centering
\caption{Detailed results for the cantilever beam with corner supports: SIMP and level-set stages.}
\label{tab:detailed_comparison-4Legs}
\begin{tabular}{p{2.8cm}p{2.0cm}|w{c}{0.9cm}w{c}{0.9cm}w{c}{0.9cm}w{c}{0.9cm}w{c}{0.9cm}|w{c}{1.0cm}}
\toprule
 &  & \multicolumn{5}{c|}{Maximum density change $\Delta\rho_{\max}$ [\%]} & \\
    Step & Metric & 0.5 & 1 & 2 & 4 & 8 & Porous \\
\midrule
1. step - SIMP         & Iterations [\si{1}] & 312 & 64 & 38 & 22 & 13 & --- \\
(extracted)  & Time [\si{\second}]   & 978.0 & 203.4 & 122.5 & 72.2 & 44.3 & --- \\
             & Compliance [\si{\joule}] & 66.95 & 68.88 & 69.10 & 69.14 & 69.26 & --- \\
             & Vol. frac. [\si{1}] & 0.400 & 0.401 & 0.401 & 0.400 & 0.400 & --- \\
\midrule
2. step - Level-set    & Iterations [\si{1}] & 19 & 32 & 32 & 32 & 19 & 151 \\
(extracted)  & Time [\si{\second}]   & 291.5 & 456.8 & 456.3 & 455.7 & 293.9 & 1561.2 \\
             & Compliance [\si{\joule}] & 66.35 & 67.25 & 67.22 & 66.74 & 67.59 & 66.45 \\
             & Vol. frac. [\si{1}] & 0.403 & 0.404 & 0.404 & 0.404 & 0.403 & 0.404 \\
\bottomrule
\end{tabular}
\end{table*}

\begin{table*}[!t]
\centering
\caption{Sequential methodology summary for the cantilever beam with corner supports. Speedup is relative to the porous initialization baseline.}
\label{tab:seq_method-4Legs}
\begin{tabular}{l| cc}
\toprule
Optimization & $t_{\mathrm{total}}$ [\si{\second}] & Speedup [\si{1}] \\
\midrule
SIMP $\rightarrow$ LS ($\Delta\rho_{\max} = 0.5\%$) & 1273.3 & 1.23 \\ 
SIMP $\rightarrow$ LS ($\Delta\rho_{\max} = 1\%$)  & 664.0 & 2.35 \\ 
SIMP $\rightarrow$ LS ($\Delta\rho_{\max} = 2\%$)  & 582.6 & 2.68 \\ 
SIMP $\rightarrow$ LS ($\Delta\rho_{\max} = 4\%$)  & 531.7 & 2.94 \\ 
SIMP $\rightarrow$ LS ($\Delta\rho_{\max} = 8\%$)  & 342.0 & 4.56 \\ 
\midrule
Porous $\rightarrow$ LS                             & 1561.2 & 1.00 \\
\bottomrule
\end{tabular}

\vspace{2mm}
\footnotesize
\noindent $t_{\mathrm{total}}$ = cumulative computation time (SIMP + extraction + level-set),\\ Speedup = $t_{\mathrm{porous}} / t_{\mathrm{total}}$.
\end{table*}

\subsection{MBB beam with symmetry conditions}\label{sub:MBB}

\begin{figure}[htbp]
    \centering
    \includegraphics[width=0.3\textwidth]{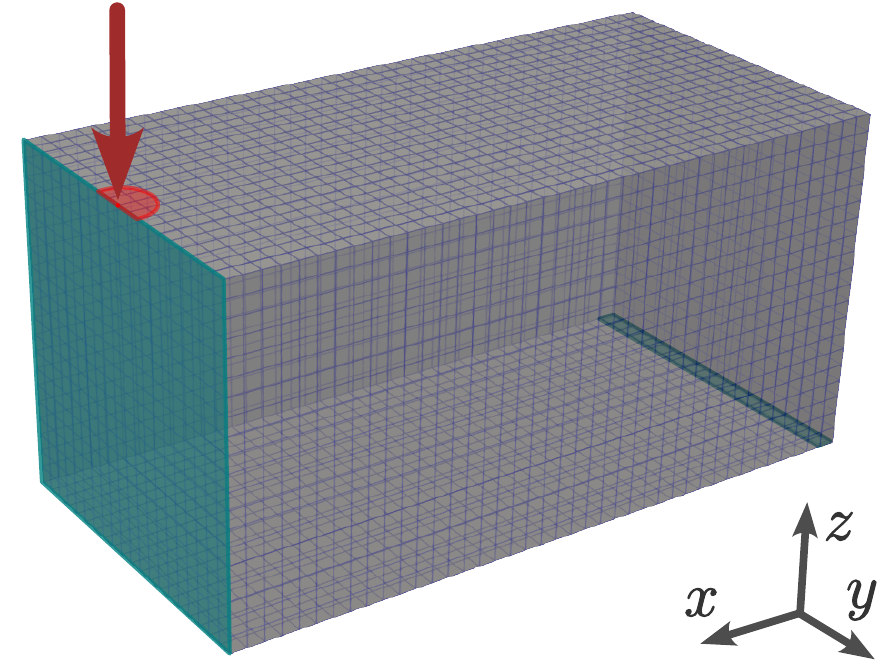}
    \caption{MBB beam problem setup: sliding boundary conditions (green) at the symmetry plane ($x = 0$) and bottom edge ($x = 2.0$); load (red) applied on a semicircular region of radius $0.1$.}
    \label{fig:MBB_geometry}
\end{figure}

\noindent The second test case considers the Messerschmitt-B\"{o}lkow-Blohm (MBB) beam, a standard benchmark in topology optimization~\cite{Sigmund2001}. Figure~\ref{fig:MBB_geometry} illustrates the problem configuration. Symmetry boundary conditions on the $x = 0$ face constrain the $x$-displacement. A simple support is applied along the bottom edge at $x = 2.0$ over a strip of width $h = 0.05$ (one element at the current resolution), constraining the $z$-displacement. A load $\boldsymbol{f} = (0, 0, -1)$~N acts on a semicircular region of radius $0.1$ centered at $(0, 0.5, 1)$ on the top face.

Tables~\ref{tab:detailed_comparison-MBB} and~\ref{tab:seq_method-MBB} present the optimization results. The SIMP iteration count follows the same trend as in the cantilever case: tighter convergence criteria require more iterations. The compliance of extracted geometries does not decrease monotonically with tighter convergence. This variability reflects the sensitivity of the iso-surface extraction to density values near the threshold $\rho_t = 0.5$: at partially-converged SIMP states, small changes in intermediate-density regions can shift which elements fall inside or outside the extracted boundary, producing compliance variations of order a few percent. This sensitivity is more pronounced in the MBB case than in the cantilever case because its symmetry-loaded topology contains thin members whose local stiffness depends sensitively on the inclusion or exclusion of individual elements near the boundary.

\begin{figure*}[!t]
    \centering
    
    \fbox{\begin{minipage}[b]{0.47\textwidth}
        \centering
        \includegraphics[width=0.48\textwidth]{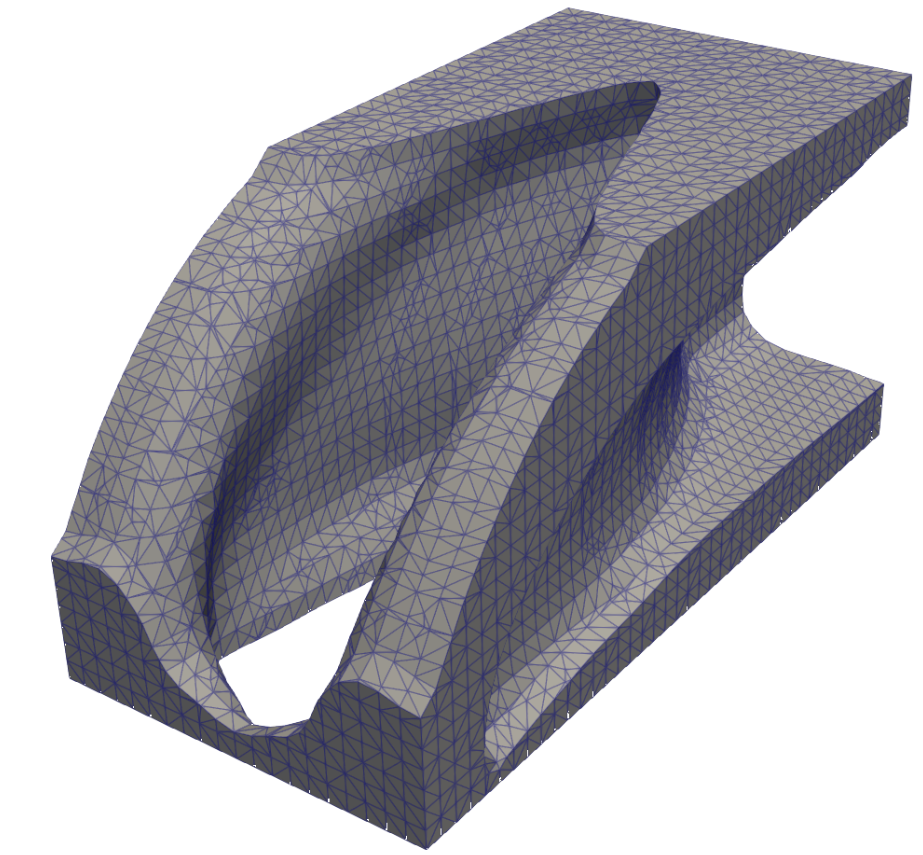}
        \hfill
        \includegraphics[width=0.48\textwidth]{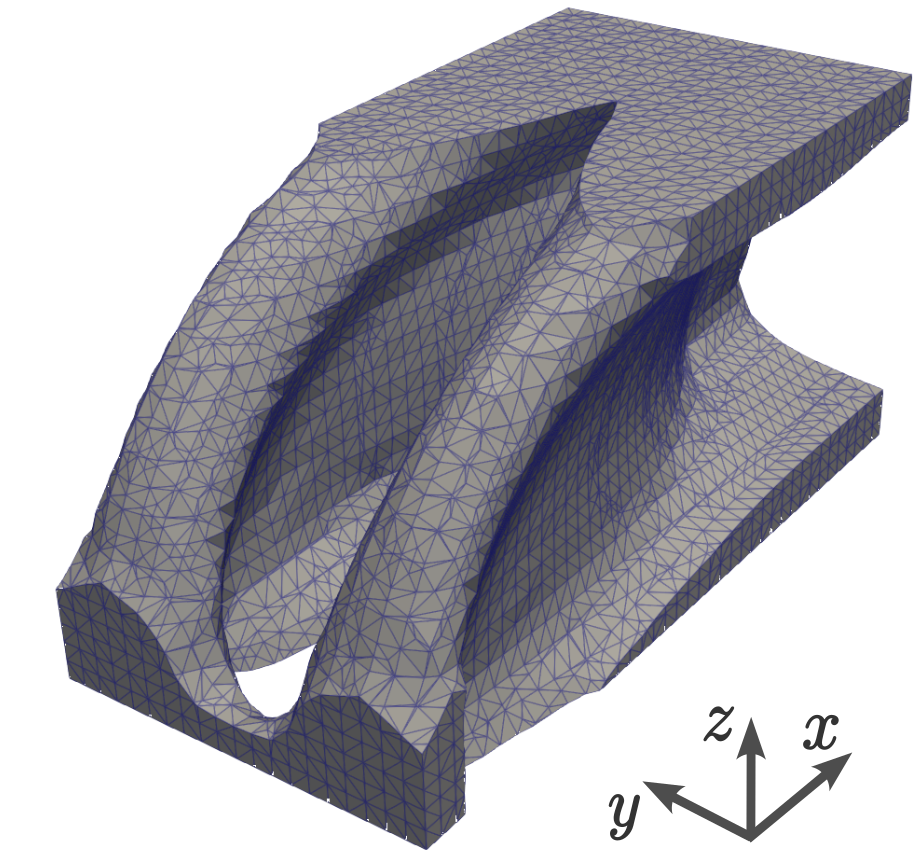}
        \\[0.1cm]
        \small\textbf{(a) $\Delta\rho_{\max} = 0.5\%$}
    \end{minipage}}
    \hfill
    \fbox{\begin{minipage}[b]{0.47\textwidth}
        \centering
        \includegraphics[width=0.48\textwidth]{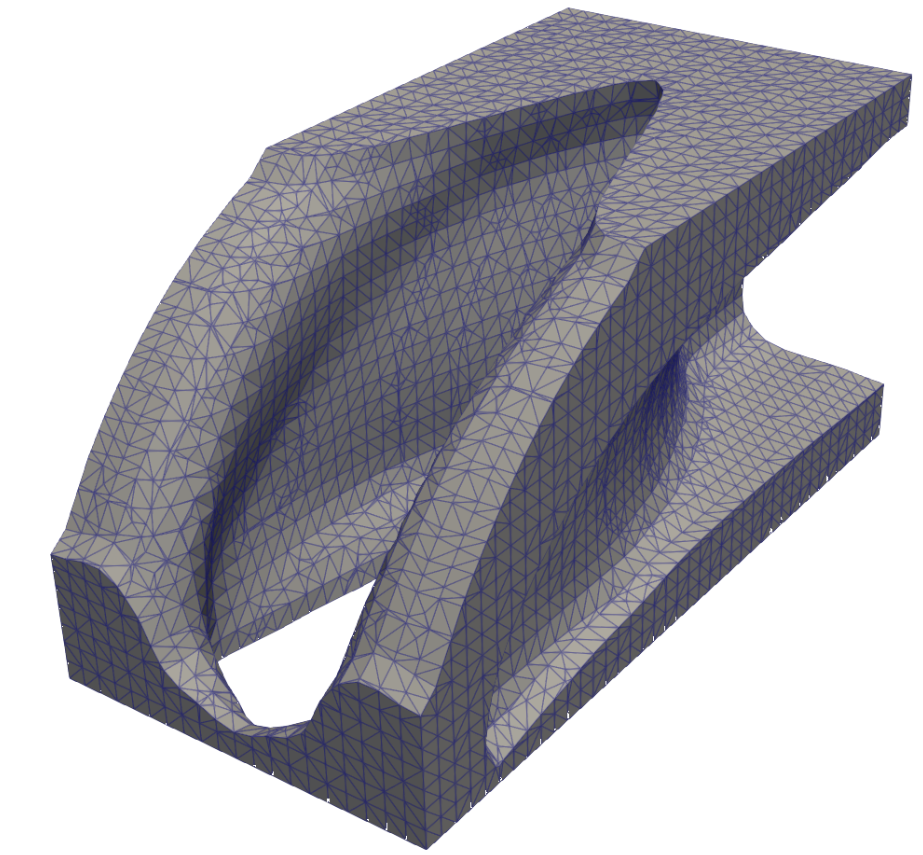}
        \hfill
        \includegraphics[width=0.48\textwidth]{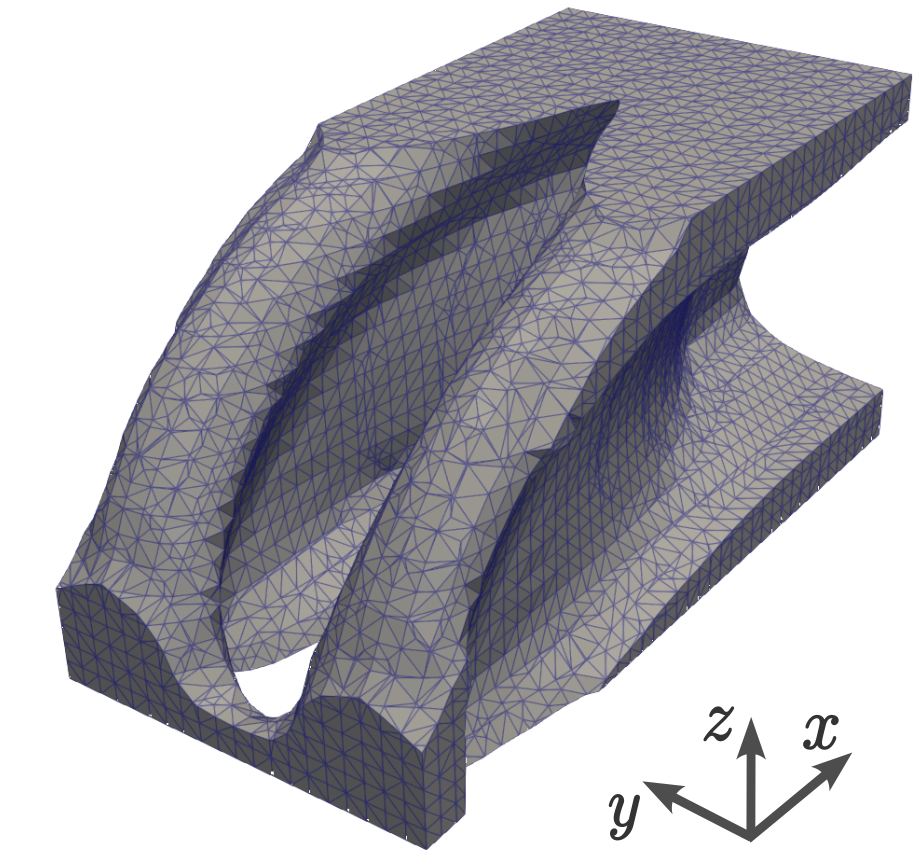}
        \\[0.1cm]
        \small\textbf{(b) $\Delta\rho_{\max} = 1\%$}
    \end{minipage}}
    
    \vspace{0.3cm}

    \fbox{\begin{minipage}[b]{0.47\textwidth}
        \centering
        \includegraphics[width=0.48\textwidth]{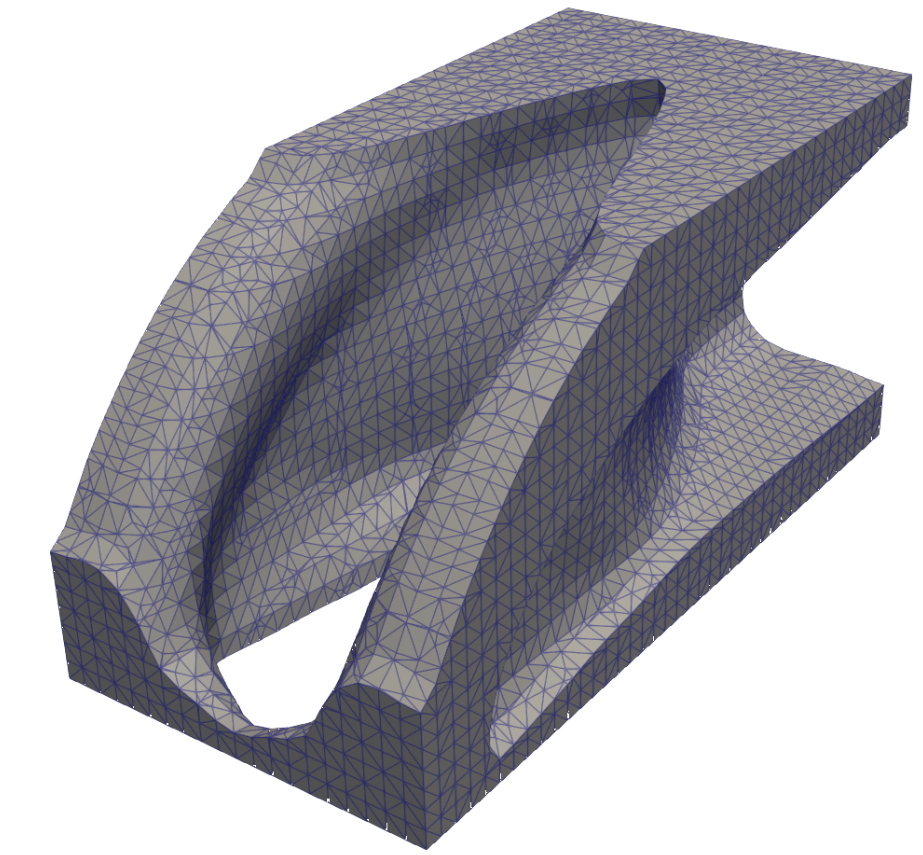}
        \hfill
        \includegraphics[width=0.48\textwidth]{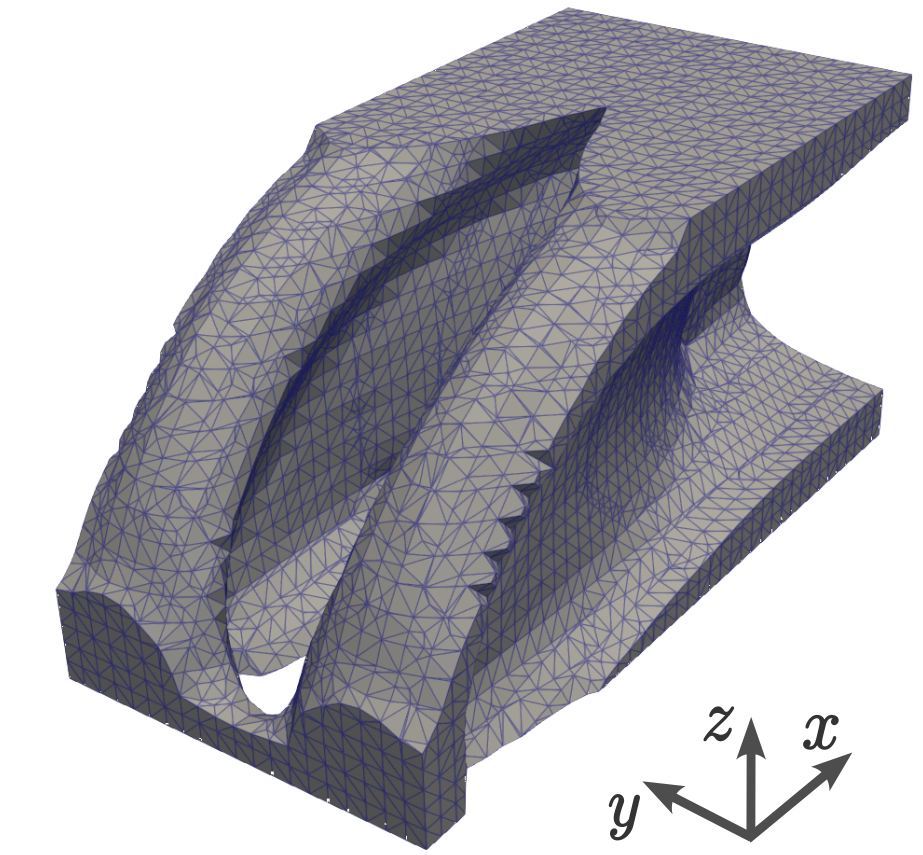}
        \\[0.1cm]
        \small\textbf{(c) $\Delta\rho_{\max} = 2\%$}
    \end{minipage}}
    \hfill
    \fbox{\begin{minipage}[b]{0.47\textwidth}
        \centering
        \includegraphics[width=0.48\textwidth]{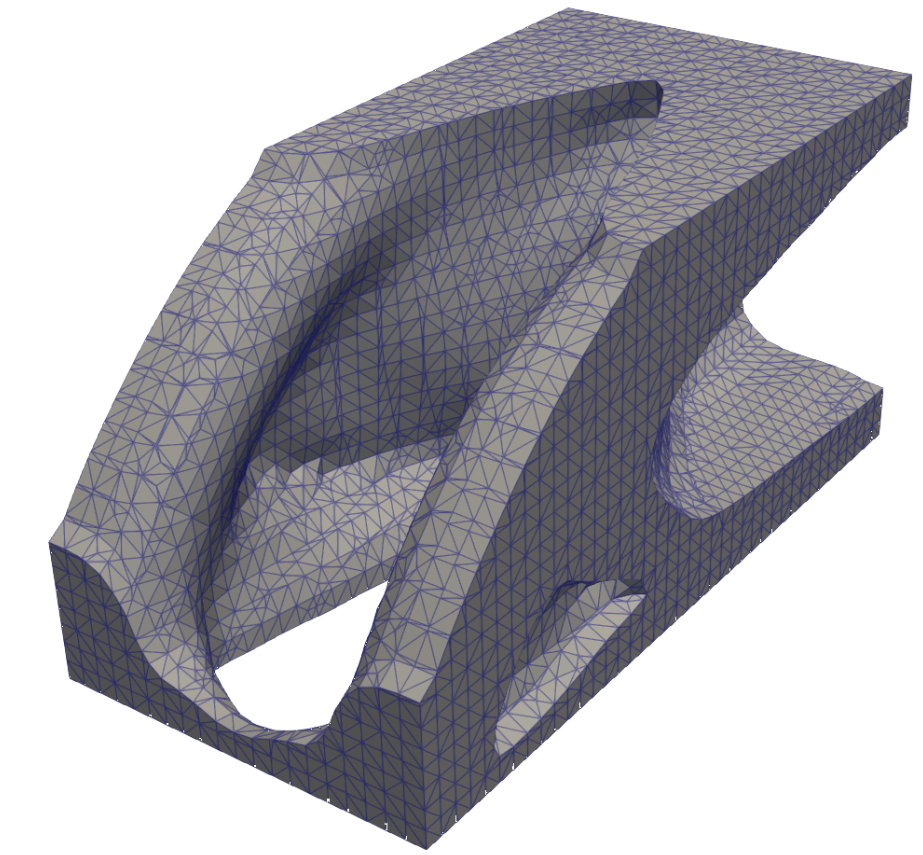}
        \hfill
        \includegraphics[width=0.48\textwidth]{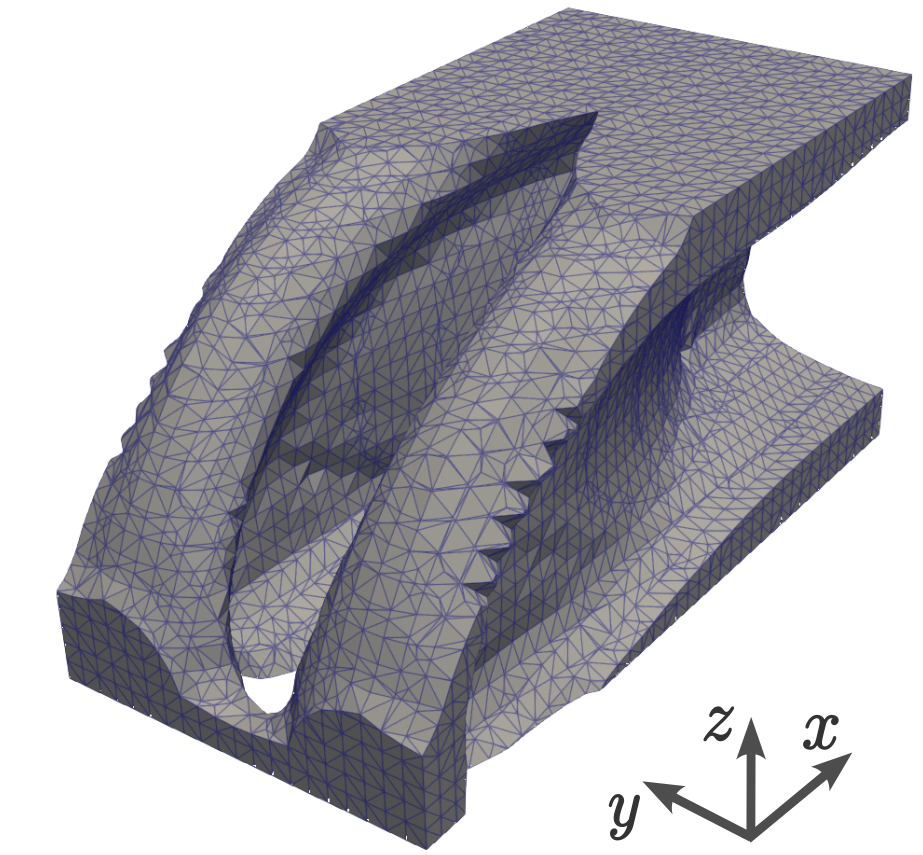}
        \\[0.1cm]
        \small\textbf{(d) $\Delta\rho_{\max} = 4\%$}
    \end{minipage}}

    \vspace{0.3cm}

    \fbox{\begin{minipage}[b]{0.47\textwidth}
        \centering
        \includegraphics[width=0.48\textwidth]{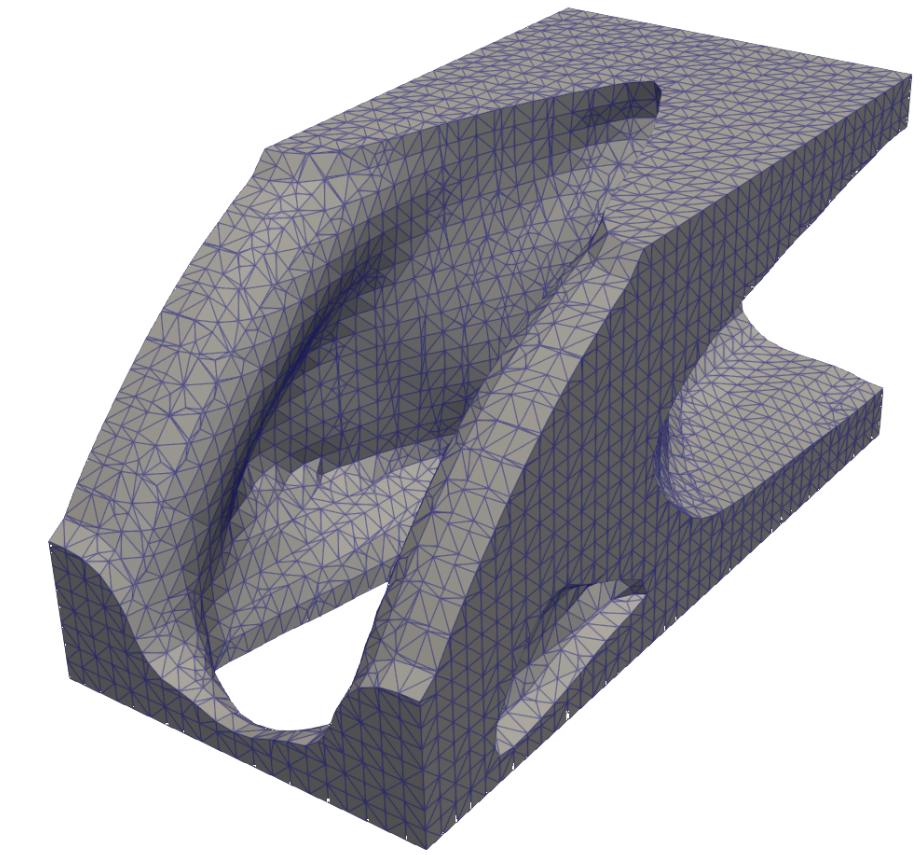}
        \hfill
        \includegraphics[width=0.48\textwidth]{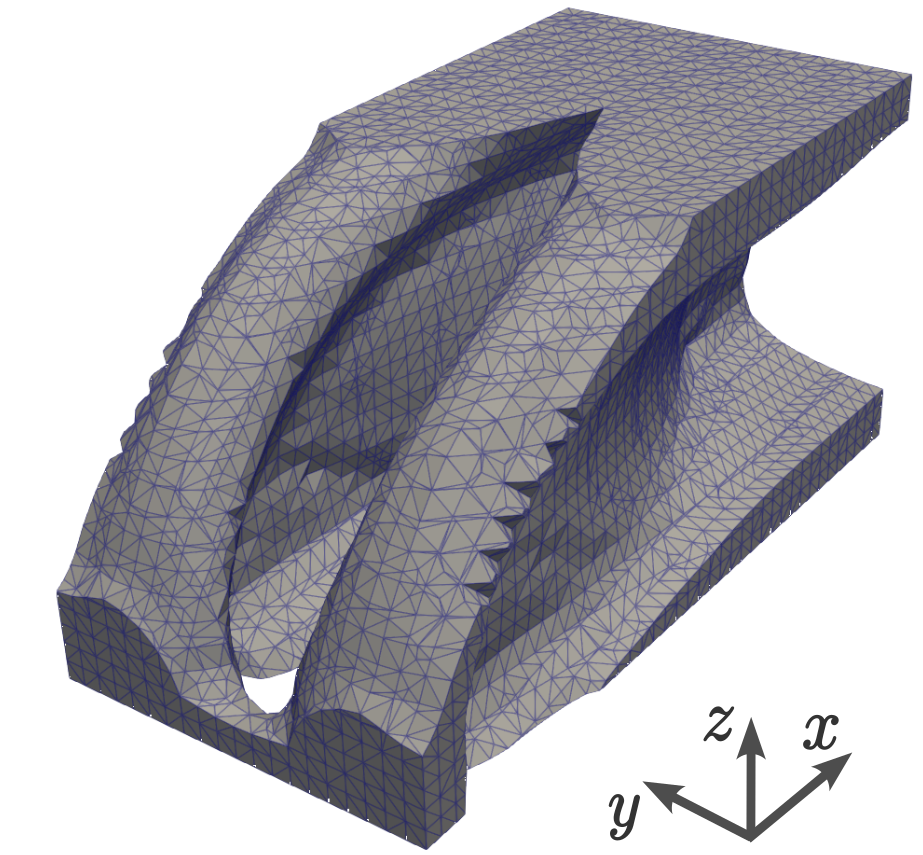}
        \\[0.1cm]
        \small\textbf{(e) $\Delta\rho_{\max} = 8\%$}
    \end{minipage}}
    \hfill
    \fbox{\begin{minipage}[b]{0.47\textwidth}
        \centering
        \includegraphics[width=0.48\textwidth]{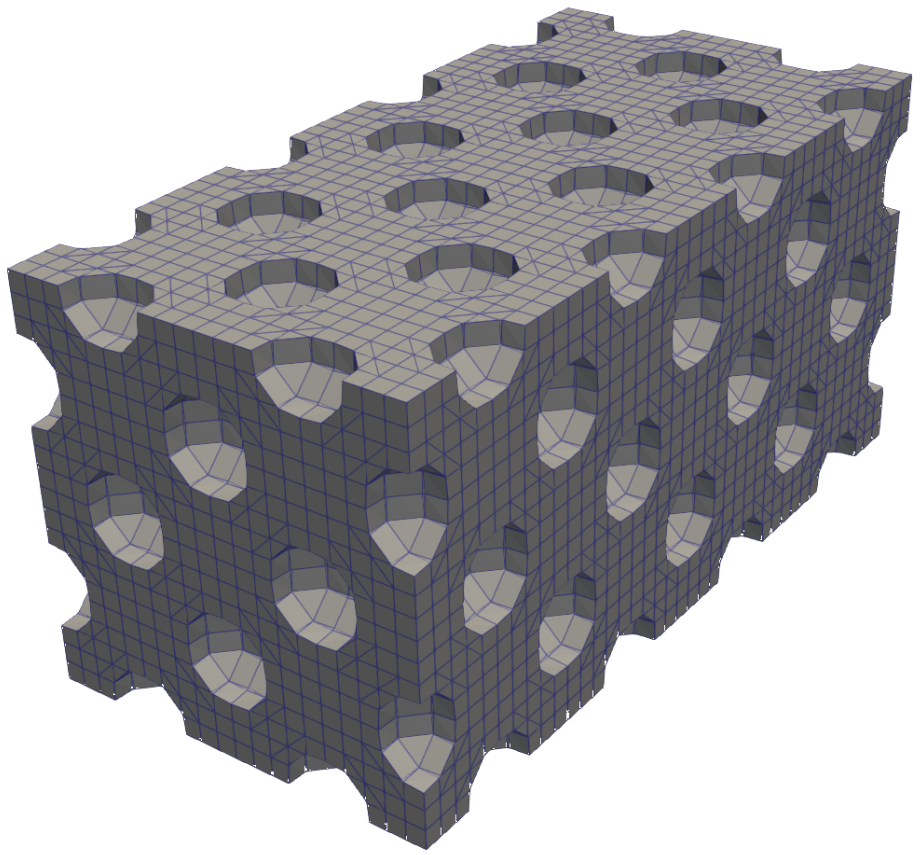}
        \hfill
        \includegraphics[width=0.48\textwidth]{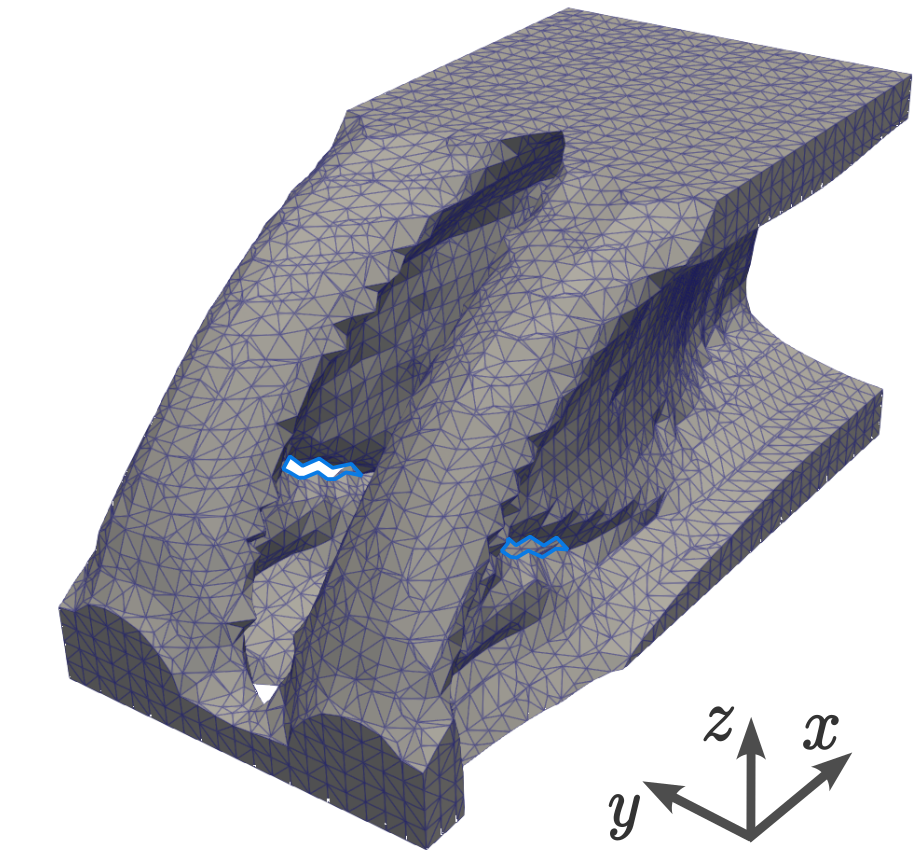}
        \\[0.1cm]
        \small\textbf{(f) Porous initialization}
    \end{minipage}}
    
    \vspace{0.2cm}
    \footnotesize Each box: SIMP extracted geometry (left) $\rightarrow$ level-set refined result (right).
    
    \caption{MBB beam optimization results. Each pair shows the extracted SIMP geometry (left) and the level-set refined result (right). Cases (a)--(e): sequential SIMP$\rightarrow$LS with varying convergence tolerance $\Delta\rho_{\max}$. Case (f): baseline from uniform porous initialization. Holes are marked to highlight topological differences between cases.}
    \label{fig:MBB_results}
\end{figure*}

The level-set stage requires fewer iterations for all sequential initializations than for the porous baseline, though the reduction is less pronounced than in the cantilever case (Table~\ref{tab:detailed_comparison-MBB}). The final compliance values after level-set optimization vary by less than $4\%$ across all configurations. Notably, the sequential initializations all converge to a compliance-superior topology (best $66.31$~J) compared to the porous baseline ($68.44$~J, a $3.2\%$ higher compliance). The porous initialization appears to fall into a suboptimal local minimum that the SIMP-informed initialization avoids (see Figure~\ref{fig:MBB_results}). On this problem the sequential framework therefore provides both a modest wall-clock speedup and a quality-improved final design.

All sequential configurations achieve speedup over the porous baseline, ranging from $1.26\times$ to $1.44\times$ (Table~\ref{tab:seq_method-MBB}). These gains are more modest than in the cantilever case because the level-set stage on the MBB problem requires more iterations regardless of initialization. The SIMP-derived starting topology, while feasible, still requires substantial boundary and topological refinement, particularly at looser SIMP convergence where a diagonal member must be removed (Figure~\ref{fig:MBB_results}d,e). The geometry extraction and SDF construction overhead is comparable to the cantilever case and negligible relative to the optimization stages.

Figure~\ref{fig:MBB_results} presents the resulting geometries. Each pair displays the extracted SIMP geometry (left) and the level-set refined result (right). Tighter SIMP convergence produces extracted geometries that progressively approach the level-set optimized shape. All sequential initializations converge to membrane-like structures without interior holes. The porous initialization yields a structure with a different topology.

\begin{figure*}[!t]
    \centering
    \begin{subfigure}[b]{0.48\textwidth}
        \centering
        \includegraphics[width=\textwidth]{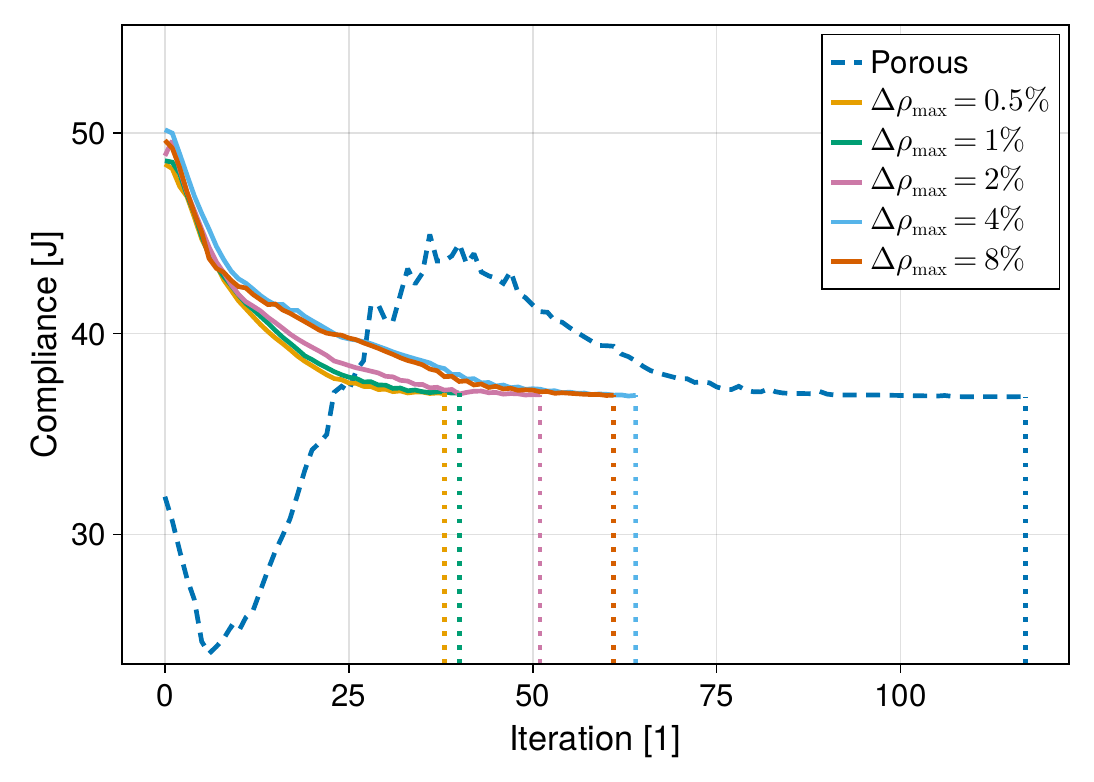}
        \caption{Compliance}
        \label{fig:MBB_conv_J}
    \end{subfigure}
    \hfill
    \begin{subfigure}[b]{0.48\textwidth}
        \centering
        \includegraphics[width=\textwidth]{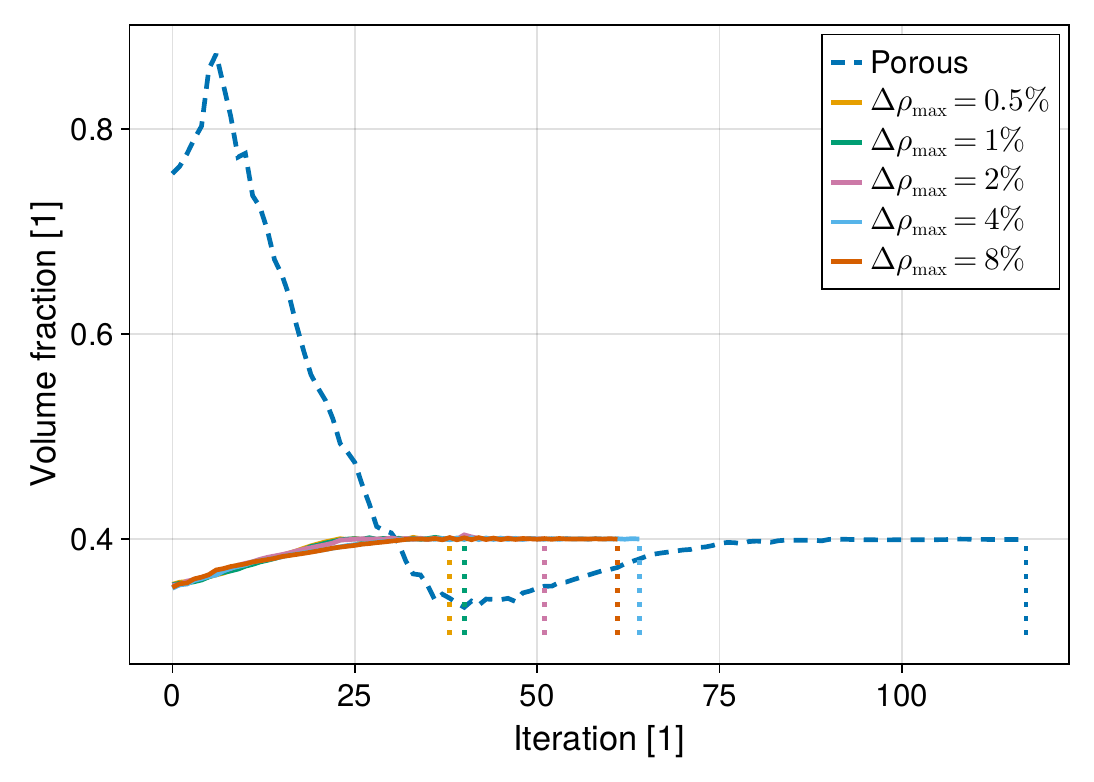}
        \caption{Volume fraction}
        \label{fig:MBB_conv_Vol}
    \end{subfigure}
    \caption{MBB beam level-set convergence histories: (a)~Compliance and (b)~volume fraction of extracted geometries. Sequential initializations (SIMP$\rightarrow$LS) with varying $\Delta\rho_{\max}$ are compared against the porous initialization baseline (dashed line).}
    \label{fig:MBB_convergence}
\end{figure*}

Figure~\ref{fig:MBB_convergence} shows the level-set convergence histories. The number of level-set iterations correlates with the extent of geometric modifications required by the level-set stage. Tighter SIMP convergence produces initializations that need only minor boundary adjustments, resulting in fewer level-set iterations. The $\Delta\rho_{\max} = 4\%$ and $8\%$ cases require more iterations because the level-set stage must remove a diagonal structural member present in the SIMP-extracted geometry (Figure~\ref{fig:MBB_results}d,e). The volume fraction plot (Figure~\ref{fig:MBB_conv_Vol}) shows that sequential initializations start close to the prescribed volume fraction, with approximately $3\%$ deviation caused by the reinitialization step. The porous initialization starts far from the target volume and requires substantially more iterations to converge.

\begin{table*}[!t]
\centering
\caption{Detailed results for the MBB beam: SIMP and level-set stages.}
\label{tab:detailed_comparison-MBB}
\begin{tabular}{p{2.8cm}p{2.0cm}|w{c}{0.9cm}w{c}{0.9cm}w{c}{0.9cm}w{c}{0.9cm}w{c}{0.9cm}|w{c}{1.0cm}}
\toprule
 &  & \multicolumn{5}{c|}{Maximum density change $\Delta\rho_{\max}$ [\%]} & \\
    Step & Metric & 0.5 & 1 & 2 & 4 & 8 & Porous \\
\midrule
1. step - SIMP         & Iterations [\si{1}] & 106 & 85 & 62 & 23 & 21 & --- \\
(extracted)  & Time [\si{\second}]   & 341.1 & 275.2 & 201.9 & 77.5 & 70.9 & --- \\
             & Compliance [\si{\joule}] & 68.92 & 68.16 & 69.96 & 68.96 & 69.95 & --- \\
             & Vol. frac. [\si{1}] & 0.405 & 0.405 & 0.405 & 0.406 & 0.406 & --- \\
\midrule
2. step - Level-set    & Iterations [\si{1}] & 38 & 40 & 51 & 64 & 61 & 117 \\
(extracted)  & Time [\si{\second}]   & 579.8 & 609.5 & 756.7 & 938.6 & 890.4 & 1280.6 \\
             & Compliance [\si{\joule}] & 66.31 & 67.04 & 67.86 & 66.53 & 68.54 & 68.44 \\
             & Vol. frac. [\si{1}] & 0.405 & 0.406 & 0.406 & 0.406 & 0.405 & 0.403 \\
\bottomrule
\end{tabular}
\end{table*}

\begin{table*}[!t]
\centering
\caption{Sequential methodology summary for the MBB beam. Speedup is relative to the porous initialization baseline.}
\label{tab:seq_method-MBB}
\begin{tabular}{l| cc}
\toprule
Optimization & $t_{\mathrm{total}}$ [\si{\second}] & Speedup [\si{1}] \\
\midrule
SIMP $\rightarrow$ LS ($\Delta\rho_{\max} = 0.5\%$) & 924.7 & 1.38 \\ 
SIMP $\rightarrow$ LS ($\Delta\rho_{\max} = 1\%$) & 888.4 & 1.44 \\ 
SIMP $\rightarrow$ LS ($\Delta\rho_{\max} = 2\%$) & 962.4 & 1.33 \\ 
SIMP $\rightarrow$ LS ($\Delta\rho_{\max} = 4\%$) & 1019.8 & 1.26 \\ 
SIMP $\rightarrow$ LS ($\Delta\rho_{\max} = 8\%$) & 965.0 & 1.33 \\ 
\midrule
Porous $\rightarrow$ LS                             & 1280.6 & 1.00 \\
\bottomrule
\end{tabular}

\vspace{2mm}
\footnotesize
\noindent $t_{\mathrm{total}}$ = cumulative computation time (SIMP + extraction + level-set),\\ Speedup = $t_{\mathrm{porous}} / t_{\mathrm{total}}$.
\end{table*}

\section{Conclusion}\label{sect:conclusion}

\noindent This paper presents a sequential topology optimization framework combining density-based and level-set methods, and evaluates its behavior on three-dimensional benchmark problems. The methodological contribution is an SDF-based geometry transfer formulated for general 3D SIMP meshes, broadening the applicability of sequential SIMP-to-level-set frameworks~\cite{Swierstra2020}. The converged SIMP density field is extracted as a signed distance function, which then initializes level-set boundary refinement. An adaptive reduction of the Hilbertian projection step size further improves convergence from near-optimal initializations. This sequential approach exploits the efficiency of density-based methods for topological changes and the geometric precision of level-set representations for boundary definition. From the SIMP perspective, the level-set stage serves as optimization-driven post-processing that refines boundary geometry beyond what purely geometric extraction can achieve.

The numerical experiments on a cantilever beam and an MBB beam illustrate the behavior of the sequential approach. In the tested configurations, the sequential approach achieves structural performance comparable to that of porous initialization baselines, with lower compliance in several cases. Tighter SIMP convergence produces extracted geometries progressively closer to the final level-set result, and all tested SIMP convergence criteria yield fewer total level-set iterations than the porous baseline. The cantilever benchmark achieves a $4.6\times$ speedup at the loosest convergence criterion, with a trade-off between speed and compliance across the tested tolerance range. Intermediate tolerances yield $2.4$--$2.9\times$ speedup at compliance comparable to the porous baseline. The MBB benchmark yields a more modest speedup ($1.26$--$1.44\times$) because its level-set stage requires more iterations regardless of initialization, and at looser tolerances additional iterations are needed to remove a diagonal member from the SIMP-extracted geometry. On this benchmark, however, the sequential initialization converges to a lower-compliance topology (best $66.31$~J) than the porous baseline ($68.44$~J, a $3.2\%$ higher compliance). This indicates that SIMP initialization can guide the level-set stage away from suboptimal local minima. The SDF-based geometry transfer introduces negligible computational overhead, requiring approximately $3.8$~s in all tested configurations.

The benchmark results suggest several practical advantages of the sequential framework over standalone level-set optimization. The SIMP stage automatically determines a starting topology, reducing the initialization sensitivity characteristic of level-set approaches and eliminating the need for dedicated hole nucleation mechanisms. The SDF representation provides a well-conditioned initialization for the Hamilton--Jacobi evolution. The level-set stage produces geometries with sharp, well-defined boundaries suitable for manufacturing, providing an optimization-based alternative to purely geometric post-processing approaches such as that presented in~\cite{Jezek2026}. A full open-source implementation of the framework is released to support reproducibility and further development (Appendix~\ref{app:data}).

This paper does not aim to provide a comprehensive efficiency study of the sequential approach. A systematic parameter study covering mesh refinement, problem-class coverage, and convergence tolerance choices is identified as future work below. The methodology was validated on two benchmark problems using ersatz material approximation on structured hexahedral meshes. The SDF transfer methodology of Section~\ref{sub:SDF} is formulated for general mesh types. The structured mesh requirement of the level-set stage stems from the specific numerical implementation (finite difference upwinding and element-wise material interpolation), not from level-set methods in general~\cite{VanDijk2013}. In the tested configurations, the Hilbertian projection method performed only limited boundary modifications, preserving the topology established by SIMP. In the MBB case with looser SIMP tolerances, the level-set stage additionally produced a topological change by removing a diagonal structural member. The extent of topology modification therefore depends on the proximity of the SIMP initialization to the level-set optimum. The level-set stage may converge to a local minimum determined by the density-based initialization. The effect of algorithmic parameters on convergence from near-optimal initializations has not been systematically studied. Alternative level-set formulations and constraint handling strategies~\cite{VanDijk2013} may be better suited to the sequential approach than the Hamilton--Jacobi evolution with Hilbertian projection used here.

Future research directions include a systematic convergence study across a broader set of problems, which is necessary to establish the generality and efficiency of the sequential approach. Another direction is the extension to unfitted finite element formulations~\cite{Wegert2025}. Such an extension would remove the structured-mesh requirement of the level-set stage and exploit the mesh-agnostic SDF transfer of Section~\ref{sub:SDF}. Moreover, unfitted formulations incur substantially higher per-iteration cost than ersatz material approaches, which would amplify the wall-clock benefit of reduced iteration counts.

\paragraph{Statement:} During the preparation of this work the authors used Claude (Anthropic's AI assistant) in order to improve writing clarity and ensure consistent academic style. After using this tool/service, the authors reviewed and edited the content as needed and take full responsibility for the content of the published article.

\section*{Acknowledgements}
\noindent This research was funded by the European Union under the project Metamaterials for thermally stressed machine components (reg. no. CZ.02.01.01/00/23\_020/0008501) and the institutional support RVO:61\-38\-89\-98.\\

\noindent This work also received support from the Grant Agency of the Czech Technical University in Prague, under grant \\ No. SGS24/123/OHK2/3T/12.

\newpage

\appendix
\section{Data and code availability} \label{app:data}
\noindent Ongoing development is hosted in a public GitHub repository~\cite{Jezek2026SeqTopOptCode}: \url{https://github.com/jezekon/2026-Jezek-SeqTopOpt}.

\noindent An archived snapshot for reproducing the results presented in this paper is available on Zenodo~\cite{Jezek2026SeqTopOptZenodo}: \url{https://doi.org/10.5281/zenodo.20024424}. The archive contains:
\begin{itemize}
    \item a step-by-step guide for the sequential optimization workflow,
    \item Julia scripts for each of the four methodology stages,
    \item intermediate data (density fields, extracted geometries, SDF representations, optimized geometries),
    \item input meshes and validation outputs for the 3D cantilever and MBB benchmarks.
\end{itemize}
\noindent The provided materials are released under the Creative Commons Attribution 4.0 International License (CC BY 4.0). The Julia packages invoked by the workflow are distributed separately under their own licenses.

\printcredits

\bibliographystyle{cas-model2-names}

\bibliography{references.bib}




\end{document}